\documentclass{rsauthor}
\usepackage{amsfonts,amsmath,amsbsy,epsfig,bbm}
\usepackage[numbers]{natbib}
\begin{document}

\title{Coupling all-atom molecular dynamics simulations of ions in water
with Brownian dynamics}

\author{\large Radek Erban}

\address{{\small Mathematical Institute, University of Oxford \\ 
Radcliffe Observatory Quarter,  Woodstock Road \\
Oxford OX2 6GG, United Kingdom \\ 
\rule{0pt}{4mm}
e-mail: erban@maths.ox.ac.uk}}

\date{\today}

\abstract{Molecular dynamics (MD) simulations of ions (K$^+$, Na$^+$,
Ca$^{2+}$ and Cl$^-$) in aqueous solutions are investigated. Water is 
described using the SPC/E model. A stochastic coarse-grained description 
for ion behaviour is presented and parameterized using MD simulations.
It is given as a system of coupled stochastic and ordinary differential
equations, describing the ion position, velocity and acceleration.
The stochastic coarse-grained model provides an intermediate 
description between all-atom MD simulations and Brownian dynamics
(BD) models. It is used to develop a multiscale method which uses 
all-atom MD simulations in parts of the computational domain and 
(less detailed) BD simulations in the remainder of the domain.
}

%\classification{...}

\maketitle
\section{Introduction}

Molecular dynamics (MD) simulations of ions in aqueous solutions 
are limited to modelling processes in relatively small domains
containing (only) several thousands of water 
molecules~\cite{Koneshan:1998:SSD,Kohagen:2014:ADC}. Ions play
important physiological functions in living cells which typically
consist of $10^{10}$--$10^{12}$ water molecules. In particular, 
processes which include transport of ions between different parts 
of a cell cannot be simulated using standard all-atom MD approaches.
Coarser models are instead used in applications. Examples include
Brownian dynamics (BD) simulations \cite{Dobramysl:2015:PMM} 
and mean-field Poisson-Nernst-Planck equations \cite{Corry:2000:TCT}.
In BD methods, individual trajectories of ions are described using
\begin{equation}
\mbox{d}X_i = \sqrt{2 D} \; \mbox{d}W_i, 
\qquad
i=1,2,3,
\label{BDSDE}
\end{equation}
where ${\mathbf X} = [X_1,X_2,X_3]$ is the position of the ion,
$D$ is its diffusion constant and $W_i$, $i=1,2,3$, are three 
independent Wiener processes~\cite{Erban:2007:PGS}. 
BD description~(\ref{BDSDE}) does not explicitly include
solvent molecules in the simulation. Moreover, in applications, 
equation~(\ref{BDSDE}) can be discretized using a
(nanosecond) time step which is much larger than the typical time 
step of MD simulations (femtosecond)~\cite{Leimkuhler:2015:MD}. 
This makes BD less computationally intensive than the 
corresponding MD simulations.

Longer time steps of BD simulations enable efficient simulations 
of ion transport between different parts of the cell, but they
limit the level of detail which can be incorporated into
the model. For example, intracellular calcium is regulated by 
the release of Ca$^{2+}$ ions from the endoplasmic reticulum 
via inisitol-4,5-triphosphate receptor (IP$_3$R) channels.
BD models in the literature use equation~(\ref{BDSDE}) to describe
trajectories of calcium ions~\cite{Dobramysl:2015:PMM,Flegg:2013:DSN}. 
The conformational changes between the open and closed states 
of IP$_3$R channels are controlled by the binding
of Ca$^{2+}$ to activating and inhibitory binding sites. BD
models postulate that binding of an ion occurs with some probability
whenever the distance between the ion and an empty site is less 
than the specific distance, the so called reaction 
radius~\cite{Erban:2009:SMR,Lipkova:2011:ABD}. Although details of the 
binding process are known~\cite{Shinohara:2011:MBB,Serysheva:2014:THS},
they cannot be incorporated into coarse BD models of calcium 
dynamics, because equation (\ref{BDSDE}) does not correctly 
describe short time behaviour of ion dynamics.

The calcium induced calcium release through IP$_3$R channels
is an example of a multiscale dynamical problem where MD simulations 
are important only in certain parts of the computational domain 
(close to an IP$_3$R channel), whilst in the remainder of the 
domain a coarser, less detailed, BD method could be used
(to describe trajectories of ions). Such multiscale problems 
cannot be simulated using MD methods, but there is potential
to design multiscale computational methods which compute
the desired information with an MD-level of resolution by
using MD and BD models in different parts of the computational 
domain~\cite{Erban:2014:MDB}.

In \cite{Erban:2014:MDB}, three relatively simple and analytically
tractable MD models are studied (describing heat bath molecules 
as point particles) with the aim of developing and analyzing 
multiscale methods which use MD simulations in parts of the 
computational domain and less detailed BD simulations in the 
remainder of the domain. In this follow up paper, the same question 
is investigated in all-atom MD simulations which use the SPC/E
model of water molecules. In order to couple MD and BD simulations, 
we need to first show that the MD model is in a suitable limit 
described by a stochastic model which does not explicitly take into 
account heat bath (water) molecules. In~\cite{Erban:2014:MDB}, this 
coarser description was given in terms of Langevin dynamics. 
Considering all-atom MD simulations, the coarser stochastic model of 
an ion is more complicated than Langevin dynamics. In this paper, 
it will be given by 
\begin{eqnarray}
\mbox{d}X_i & = & V_i \; \mbox{d}t, 
\label{BDXeqAAA}
\\
\mbox{d}V_i & = & U_i \, \mbox{d}t, 
\label{BDVeqAAA}
\\
\mbox{d}U_i & = & 
(- \eta_1 \, V_i + Z_i ) \, \mbox{d}t,
\label{BDUeqAAA}
\\
\mbox{d}Z_i & = & 
- (\eta_2 \, Z_i + \eta_3 \, U_i) \, \mbox{d}t
+ 
\eta_4
\; \mbox{d}W_i, 
\qquad \quad i=1,2,3,
\label{BDZeqAAA}
\end{eqnarray}
where ${\mathbf X} \equiv [X_1,X_2,X_3]$ is the position of the ion,
${\mathbf V} \equiv [V_1,V_2,V_3]$ is its velocity,
${\mathbf U} \equiv [U_1,U_2,U_3]$ is its acceleration, 
${\mathbf Z} \equiv [Z_1,Z_2,Z_3]$ is an auxiliary variable,
$\mbox{d}{\mathbf W} \equiv [\mbox{d}W_1,\mbox{d}W_2,\mbox{d}W_3]$ 
is white noise and $\eta_j$, $j=1,2,3,4$,
are parameters. These parameters will be chosen according to 
all-atom MD simulations as discussed in Section~\ref{coarseion}. 
In Section \ref{secaccuracy}, we show that (\ref{BDXeqAAA})--(\ref{BDZeqAAA})
provides a good approximation of ion behaviour.
In Section \ref{seccoupleBDXVUZ}, we further analyse the system 
(\ref{BDXeqAAA})--(\ref{BDZeqAAA}) and show how parameters
$\eta_j$, $j=1,2,3,4$, can be connected with diffusion constant $D$
used in the BD model~(\ref{BDSDE}). 

The coarse-grained model~(\ref{BDXeqAAA})--(\ref{BDZeqAAA}) is used
as an intermediate model between the all-atom MD model and
BD description~(\ref{BDSDE}). In Section~\ref{seccoupleBDXVUZ},
we show how it can be coupled with the BD model which uses a much
larger time step than the MD model. In Section~\ref{seccoupleMDBD},
the coarse-grained model~(\ref{BDXeqAAA})--(\ref{BDZeqAAA}) is 
coupled with all-atom MD simulations. We then show that all-atom MD 
models of ions can be coupled with BD description~(\ref{BDSDE}) 
using the intermediate coarse-grained 
model~(\ref{BDXeqAAA})--(\ref{BDZeqAAA}). We conclude by discussing 
related methods developed in the literature in Section~\ref{secdiscussion}.

\section{Molecular dynamics simulations of ions in SPC/E water}
\label{MDsim}

There have been several MD models of liquid water developed in the
literature. The simplest models (for example, SPC~\cite{Berendsen:1981:IMW}, 
SPC/E~\cite{Berendsen:1987:MTE} and TIP3P~\cite{Jorgensen:1983:CSP})
include three sites in total, two hydrogen atoms and an oxygen atom. 
More complicated water models include four, five or six
sites~\cite{Huggins:2012:CLW,Mark:2001:SDT}.
In this paper, we use the three-site SPC/E model of water which
was previously used for MD simulations of
ions in aqueous solutions~\cite{Lee:1996:MDS,Koneshan:1998:SSD}.
In the SPC/E model, the charges ($q_h = 0.4238\,$e)
on hydrogen sites are at 1\AA{} from the Lennard-Jones center at 
the oxygen site which has negative charge $q_o = -0.8476\,$e. 
The HOH angle is 109.47$^\circ$. We use the RATTLE 
algorithm~\cite{Andersen:1983:RVV} to satisfy constraints 
between atoms of the same water molecule. 

We investigate four ions (K$^+$, Na$^+$, Ca$^{2+}$ and Cl$^-$) at 
25$\,^\circ$C using MD parameters presented in~\cite{Lee:1996:MDS}.
Let us consider a water molecule and let us denote 
by $r_{i0}$ (resp., $r_{i1}$ and $r_{i2}$) the distance 
between the ion and the oxygen site (resp., the first 
and second hydrogen sites). The pair potential between the water 
molecule and the ion is then given 
by~\cite{Koneshan:1998:SSD,Lee:1996:MDS}, 
\begin{equation}
A_{io} \left( \frac{1}{r_{io}} \right)^{12}
-
B_{io} \left( \frac{1}{r_{io}} \right)^{6}
+
k_e \, \frac{q_i q_o}{r_{io}}
+
k_e \, \frac{q_i q_h}{r_{i1}}
+
k_e \, \frac{q_i q_h}{r_{i2}},
\label{pairpotMD}
\end{equation}
where $A_{io}$ and $B_{io}$ are Lennard-Jones parameters between 
the oxygen on the water molecule and the ion, 
$k_e$ is Coulomb's constant and $q_i$ is the charge on the ion. The 
values of parameters are given for four ions considered in 
Table~\ref{tableMDparam}.%
\begin{table}
\centerline{
\begin{tabular}{|c||c|c|c|c|}
\hline 
\rule{0pt}{4.8mm}
& 
\raise 0.681mm \hbox{$A_{io}$} 
& 
\raise 0.681mm \hbox{$B_{io}$} 
& 
\raise 0.681mm \hbox{$q_i$} 
& 
\raise 0.681mm \hbox{$M$} 
\\
\rule{0pt}{4.8mm}
\raise 0.681mm \hbox{ion} 
&
\raise 0.681mm \hbox{[Da \AA$^{14}$ ps$^{-2}$]} 
& 
\raise 0.681mm \hbox{[Da \AA$^{8}$ ps$^{-2}$]} 
& 
\raise 0.681mm \hbox{[e]} 
& 
\raise 0.681mm \hbox{[Da]} 
\\
\hline 
\hline 
\rule{0pt}{4.8mm}
K$^+$ & $2.8973 \times 10^8$ & $2.4587 \times 10^5$ & +1 & 39.0983 \\
\hline 
\rule{0pt}{4.8mm}
Na$^+$ & $6.6813 \times 10^7$ & $1.1807 \times 10^5$ & +1 & 22.9898 \\
\hline 
\rule{0pt}{4.8mm}
Ca$^{2+}$ & $1.1961  \times 10^8$& $1.5797 \times 10^5$ & +2 & 40.078 \\
\hline 
\rule{0pt}{4.8mm}
Cl$^-$ & $1.8038 \times 10^9$ & $6.1347 \times 10^5$ & -1 &  35.453 \\
\hline
\end{tabular}
}
\caption{{\it Parameters of all-atom MD simulations of ions.}
\label{tableMDparam}}
\end{table}
We express mass in daltons (Da), 
length in \aa{}ngstr\"oms (\AA) and time in picoseconds
(ps), consistently in the whole paper. Using these units, 
the parameters of the Lennard-Jones potential between 
the oxygen sites on two SPC/E water molecules are 
$A_{oo}=2.6334 \times 10^8$ Da \AA$^{14}$ ps$^{-2}$
and $B_{oo}=2.6171 \times 10^5$ Da \AA$^{8}$ ps$^{-2}$.

We consider a cube of side $L=24.83\,$\AA{} containing 511 water molecules
and 1 ion, i.e. we have $8^3=512$ molecules in our simulation box.
In the following section, we use standard NVT simulations where the
temperature is controlled using Nos\'e-Hoover 
thermostat~\cite{Nose:1984:UFC,Hoover:1985:CDE} and the
number of particles is kept constant by implementing periodic boundary
conditions. In particular, we assume that our simulation box is 
surrounded by periodic copies of itself. Then the long-range 
(Coulombic) interactions can be computed using several different
approaches, including the Ewald summation or the reaction field 
method~\cite{Perera:1995:ETL,Nymand:2000:ESR}. We use the cutoff
sphere of radius $L/2$ and the reaction field correction as implemented
in~\cite{Koneshan:1998:SSD}. This approach is more suitable for 
multiscale methods (studied later in Section \ref{seccoupleMDBD})
than the Ewald summation technique. The MD timestep is for all 
MD simulations in this paper chosen as $\Delta t = 10^{-3}$ ps $=1$ fs.

\section{Parametrization of the coarse-grained model of ion}
\label{coarseion}

In MD simulations, an ion is descibed by its position 
${\mathbf X} \equiv [X_1,X_2,X_3]$ and 
velocity ${\mathbf V} \equiv [V_1,V_2,V_3]$ which
evolve according to 
\begin{eqnarray}
\mbox{d}X_i & = & V_i \; \mbox{d}t, 
\label{BDXeqMD}
\\
M \mbox{d}V_i & = & F_i \, \mbox{d}t, 
\qquad \quad
i=1,2,3,
\label{BDVeqMD}
\end{eqnarray}
where $M$ is the mass of the ion (given in Table~\ref{tableMDparam})
and ${\mathbf F} \equiv [F_1,F_2,F_3]$ is the force acting on the ion. 
We use all-atom MD simulations as described in Section~\ref{MDsim} 
to estimate diffusion coefficient $D$ and second moments of $V_i$ 
and $U_i = F_i/M$, $i=1,2,3$. They are given for four ions considered 
in Table \ref{tableMDaverages}.
\begin{table}
\centerline{
\begin{tabular}{|c||c|c|c|c|}
\hline 
\rule{0pt}{4.8mm}
& 
\raise 0.681mm \hbox{$D$} 
& 
\raise 0.681mm \hbox{$\langle V_i^2 \rangle$} 
& 
\raise 0.681mm \hbox{$\langle U_i^2 \rangle$} 
& 
\raise 0.681mm \hbox{$\langle Z_i^2 \rangle$} 
\\
\rule{0pt}{4.8mm}
\raise 0.681mm \hbox{ion} 
& \raise 0.681mm \hbox{[\AA$^2$ ps$^{-1}$]} 
& \raise 0.681mm \hbox{[\AA$^2$ ps$^{-2}$]} 
& \raise 0.681mm \hbox{[\AA$^2$ ps$^{-4}$]} 
& \raise 0.681mm \hbox{[\AA$^2$ ps$^{-6}$]} 
\\
\hline 
\hline 
\rule{0pt}{4.8mm}
K$^+$ & 0.183 & 6.32 & $4.86 \times 10^3$ & $1.65 \times 10^7$ \\
\hline 
\rule{0pt}{4.8mm}
Na$^+$ & 0.128 & 10.8 & $2.21 \times 10^4$ & $8.88 \times 10^7$ \\
\hline 
\rule{0pt}{4.8mm}
Ca$^{2+}$ & 0.053 & 6.18 & $1.87 \times 10^4$ & $9.23 \times 10^7$ \\
\hline 
\rule{0pt}{4.8mm}
Cl$^-$ & 0.177 & 6.98 & $6.56 \times 10^3$ &  $2.97 \times 10^7$ \\
\hline
\end{tabular}
}
\caption{{\it Average values obtained by all-atom MD simulations.} 
\label{tableMDaverages}}
\end{table}
To estimate $\langle U_i^2 \rangle$, we calculate the average force
in the $i$-th direction $\langle F_i^2 \rangle$ where 
$\langle \cdot\rangle$ denotes an average over sufficiently large
time interval (nanosecond) of MD simulations. Taking into account 
the symmetry of the problem, we estimate 
$\langle U_i^2 \rangle = \langle F_i^2 \rangle / M^2$
as the average over all three dimensions
$$
\frac{\langle U_1^2 \rangle  
+ 
\langle U_2^2 \rangle 
+ 
\langle U_3^2 \rangle}{3}.
$$
This value is reported in Table~\ref{tableMDaverages}. In the same
way, the reported values of $\langle V_i^2 \rangle$ are computed
as averages over all three dimensions. Diffusion constant $D$ can be
estimated by calculating mean square displacements or velocity
autocorrelation functions. In Table~\ref{tableMDaverages}, we
report the values of $D$ which were estimated in~\cite{Koneshan:1998:SSD} 
by calculating mean square displacements.

Let us consider the coarse-grained model~(\ref{BDXeqAAA})--(\ref{BDZeqAAA})
and let $\langle \cdot\rangle$ denotes an average over many realizations
of a stochastic process.  Multiplying equations
(\ref{BDVeqAAA}) and (\ref{BDUeqAAA}) by $V_i$ and $U_i$,
respectively, we obtain the following ODEs for second moments:
\begin{eqnarray}
\frac{\mbox{d}}{\mbox{d}t}
\langle V_i^2 \rangle & = & 
2 \, \langle U_i V_i \rangle, 
\label{V2eq}
\\
\frac{\mbox{d}}{\mbox{d}t}
\langle U_i^2 \rangle
& = &
- 
2 \eta_1 \langle U_i V_i \rangle
+ 
2 \langle U_i Z_i \rangle.
\label{U2eq}
\end{eqnarray}
Consequently, we obtain that $\langle U_i V_i \rangle = 0$ and 
$\langle U_i Z_i \rangle = 0$ at steady state. Multiplying equations
(\ref{BDVeqAAA})--(\ref{BDZeqAAA}) by $V_i$, $U_i$ and $Z_i$, 
and taking averages, we obtain
\begin{eqnarray}
\frac{\mbox{d}}{\mbox{d}t}
\langle U_i V_i \rangle
& = &
\langle U_i^2 \rangle
-
\eta_1 \, \langle V_i^2 \rangle 
+ 
\langle V_i Z_i \rangle, 
\label{UVeq}
\\
\frac{\mbox{d}}{\mbox{d}t}
\langle V_i Z_i \rangle
& = & 
\langle U_i Z_i\rangle
-
\eta_2 \langle V_i Z_i \rangle
-
\eta_3 \langle U_i V_i \rangle.
\label{VZeq}
\end{eqnarray}
Using $\langle U_i V_i \rangle = 0$ and $\langle U_i Z_i \rangle = 0$,
we obtain that $\langle V_i Z_i \rangle = 0$ at steady state and
\begin{equation}
\eta_1
=
\frac{\langle U_i^2 \rangle}{\langle V_i^2 \rangle}. 
\label{eta1estimate}
\end{equation}
This equation is used in Table~\ref{tableetavalues} to estimate
$\eta_1$ using the MD averages $\langle U_i^2 \rangle$ and
$\langle V_i^2 \rangle$ which are given in Table~\ref{tableMDaverages}.
\begin{table}
\centerline{
\begin{tabular}{|c||c|c|c|c|}
\hline 
\rule{0pt}{4.8mm}
& 
\raise 0.681mm \hbox{$\eta_1$} 
& 
\raise 0.681mm \hbox{$\eta_2$} 
& 
\raise 0.681mm \hbox{$\eta_3$} 
& 
\raise 0.681mm \hbox{$\eta_4$} 
\\
\rule{0pt}{4.8mm}
\raise 0.681mm \hbox{ion} 
& \raise 0.681mm \hbox{[ps$^{-2}$]} 
& \raise 0.681mm \hbox{[ps$^{-1}$]} 
& \raise 0.681mm \hbox{[ps$^{-2}$]} 
& \raise 0.681mm \hbox{[\AA \, ps$^{-7/2}$]} 
\\
\hline 
\hline 
\rule{0pt}{4.8mm}
K$^+$ & 768.7 & 152.5 & $3.393 \times 10^3$ & $7.094 \times 10^{4}$ \\
\hline 
\rule{0pt}{4.8mm}
Na$^+$ & $2.044 \times 10^3$ & 166.1 & $4.020 \times 10^3$ 
& $1.717 \times 10^{5}$ \\
\hline 
\rule{0pt}{4.8mm}
Ca$^{2+}$ & $3.026 \times 10^3$ & 190.2 & $4.933 \times 10^3$ 
& $1.874 \times 10^{5}$ \\
\hline 
\rule{0pt}{4.8mm}
Cl$^-$ & 940.0 & 189.7 & $4.524 \times 10^3$ & $1.061 \times 10^{5}$ \\
\hline
\end{tabular}
}
\caption{{\it Values of $\eta_j$, $j=1,2,3,4$,
estimated using all-atom MD simulations. \label{tableetavalues}}}
\end{table}
Since we know the value of $\eta_1$, we can also estimate the value of
$\langle Z_i^2 \rangle$ by calculating the second moment
of
\begin{equation}
\langle Z_i^2 \rangle
\approx
\left\langle 
\left( 
\frac{U_i(t+\Delta t)-U_i(t)}{\Delta t}
+
\eta_1 \, V_i
\right)^2
\right\rangle.
\end{equation}
This value is reported in the last column of
Table~\ref{tableMDaverages}. 
Multiplying equation
(\ref{BDUeqAAA}) by $Z_i$ and equation (\ref{BDZeqAAA}) by $U_i$,
we obtain
\begin{equation}
\frac{\mbox{d}}{\mbox{d}t}
\langle U_i Z_i \rangle
 =  
\langle Z_i^2 \rangle
-
\eta_1 \langle V_i Z_i \rangle
-
\eta_2 \langle U_i Z_i \rangle
- 
\eta_3 \langle U_i^2 \rangle.
\label{UZeq}
\end{equation}
Using $\langle U_i Z_i \rangle = 0$ and $\langle V_i Z_i \rangle = 0$,
we obtain at steady state
\begin{equation}
\eta_3
=
\frac{\langle Z_i^2 \rangle}{\langle U_i^2 \rangle}.
\label{eta3estimate}
\end{equation}
Multiplying equation (\ref{BDXeqAAA}) by $X_i$, $V_i$, $U_i$ and $Z_i$
and equations (\ref{BDVeqAAA})--(\ref{BDZeqAAA}) by $X_i$ 
and taking averages, we obtain the following 
system of ODEs for second moments:
\begin{eqnarray}
\frac{\mbox{d}}{\mbox{d}t}
\langle X_i^2 \rangle & = & 
2 \langle X_i V_i \rangle,
\label{X2eq}
\\
\frac{\mbox{d}}{\mbox{d}t}
\langle X_i V_i \rangle
& = & 
\langle V_i^2 \rangle
+
\langle X_i U_i \rangle,
\label{XVeq}
\\
\frac{\mbox{d}}{\mbox{d}t}
\langle X_i U_i \rangle
& = & 
\langle U_i V_i \rangle
-
\eta_1 \langle X_i V_i \rangle
+
\langle X_i Z_i \rangle,
\label{XUeq}
\\
\frac{\mbox{d}}{\mbox{d}t}
\langle X_i Z_i \rangle
& = & 
\langle V_i Z_i\rangle
-
\eta_2 \langle X_i Z_i \rangle
-
\eta_3 \langle X_i U_i \rangle.
\label{XZeq}
\end{eqnarray}
Consequently, we obtain at steady state
$\langle X_i V_i \rangle = D$, 
$\langle X_i U_i  \rangle
= 
- \langle V_i^2  \rangle,
$
$\langle X_i Z_i  \rangle
= 
\eta_1 D
$ 
and
$$
\eta_2
=  
- \frac{\eta_3 \langle X_i U_i \rangle}{\langle X_i Z_i \rangle}
=
\frac{\eta_3 \langle V_i^2  \rangle}{\eta_1 D}.
$$
Using (\ref{eta1estimate}) and (\ref{eta3estimate}), we have
\begin{equation}
\eta_2
=
\frac{\langle Z_i^2 \rangle}{D}
\left(
\frac{\langle V_i^2 \rangle}{\langle U_i^2 \rangle}
\right)^2. 
\label{eta2estimate}
\end{equation}
Finally, multiplying equation (\ref{BDZeqAAA}) by $Z_i$, we obtain
\begin{equation}
\frac{\mbox{d}}{\mbox{d}t}
\langle Z_i^2 \rangle
=
- 
2 \eta_2 \langle Z_i^2 \rangle
- 
2 \eta_3 \langle U_i Z_i \rangle
+
\eta_4^2. 
\label{Z2eq}
\end{equation}
Consequently, we obtain at steady state
$$
\eta_4^2
= 
2 \eta_2 \langle Z_i^2 \rangle. 
$$
Using (\ref{eta2estimate}), we get
\begin{equation}
\eta_4
=
\sqrt{\frac{2}{D}}
\frac{\langle V_i^2 \rangle \langle Z_i^2 \rangle}{\langle U_i^2 \rangle}. 
\label{eta4estimate}
\end{equation}
The values calculated by (\ref{eta3estimate}), (\ref{eta2estimate})
and (\ref{eta4estimate}) are presented in Table~\ref{tableetavalues}.

\section{Accuracy of the coarse-grained model of ion}
\label{secaccuracy}

\noindent
The coarse-grained model~(\ref{BDXeqAAA})--(\ref{BDZeqAAA}) has 
four parameters $\eta_i$, $i=1,2,3,4$. To parameterize this model, 
we have used four quantities estimated from detailed MD simulations, 
diffusion constant $D$ and steady state values of 
$\langle V_i^2 \rangle$, $\langle U_i^2 \rangle$ and 
$\langle Z_i^2 \rangle$. In particular, the coarse-grained 
model~(\ref{BDXeqAAA})--(\ref{BDZeqAAA}) will give the same values
of these four quantities, including the value of diffusion constant $D$
which is the sole parameter of the BD model~(\ref{BDSDE}).
In this section, we explain why the coarse-grained description
given by~(\ref{BDXeqAAA})--(\ref{BDZeqAAA}) can be used as
an intermediate model to couple BD and MD models.

We begin by illustrating why Langevin dynamics (which is used
in~\cite{Erban:2014:MDB} for a similar multiscale problem) is
not suitable for all-atom MD simulations studied in this paper.
In~\cite{Erban:2014:MDB}, a few (heavy) particles with mass $M$ and 
radius $R$ are considered in the heat bath consisting of a 
large number of light point particles with masses $m \ll M$. 
The collisions of particles are without friction, which means 
that post-collision velocities can be computed using the conservation 
of momentum and energy. In this case, it can be shown that the 
description of heavy particles converges in an apropriate limit to 
Brownian motion given by equation~(\ref{BDSDE}). One can also 
show that the model converges to Langevin dynamics (in the limit 
$m/M \to 0$) \cite{Holley:1971:MHP,Durr:1981:MMB,Dunkel:2006:RBM}:
\begin{eqnarray}
\mbox{d}X_i & = & V_i \; \mbox{d}t, 
\label{BDXeq}
\\
\mbox{d}V_i & = & - \gamma \, V_i \, \mbox{d}t 
+ \gamma \sqrt{2 D} \; \mbox{d}W_i, 
\quad
i=1,2,3,
\label{BDVeq}
\end{eqnarray}
where ${\mathbf X} \equiv [X_1,X_2,X_3]$ is the position of a diffusing
molecule, ${\mathbf V} \equiv [V_1,V_2,V_3]$ is its velocity, $D$ is
the diffusion coefficient and $\gamma$ is the friction coefficient.
In~\cite{Erban:2014:MDB}, Langevin dynamics~(\ref{BDXeq})--(\ref{BDVeq})
is used as an intermediate model which enables the implementation of 
BD description~(\ref{BDSDE}) and the original detailed model
in different parts of the computational domain. 

Langevin dynamics~(\ref{BDXeq})--(\ref{BDVeq}) describes a diffusing 
particle in terms of its position and velocity, i.e. it uses the same
independent variables for the description of an ion as the MD
model~(\ref{BDXeqMD})--(\ref{BDVeqMD}). Langevin dynamics can
be further reduced to BD model~(\ref{BDSDE}) in the overdamped 
limit $\gamma \to \infty$. However, it cannot be used 
as an intermediate model between BD and all-atom MD simulations 
considered in this paper, because it does not correctly describe 
the ion behaviour at times comparable to the MD timestep $\Delta t$.
To illustrate this, let us parameterize Langevin
dynamics~(\ref{BDXeq})--(\ref{BDVeq}) using diffusion constant 
$D$ and the second velocity moment $\langle V_i^2 \rangle$ estimated 
from all-atom MD simulations. To get the same second moment of 
velocity, Langevin dynamics requires that we choose 
\begin{equation}
\gamma = \frac{\langle V_i^2 \rangle}{D}.
\label{gammaeq}
\end{equation} 
Discretizing equation~(\ref{BDVeq}), the ion acceleration during 
one time step is
\begin{equation}
\frac{V_i(t+\Delta t) - V_i(t)}{\Delta t} 
= 
- \gamma \, V_i(t) 
+ \gamma \sqrt{\frac{2 D}{\Delta t}} \; \xi_i
\label{LDaccel}
\end{equation} 
where $[\xi_{1},\xi_{2},\xi_{3}]$ is a vector of normally 
distributed random numbers with zero mean and unit variance.
Using~(\ref{gammaeq}), the second moment of the right hand side
of~(\ref{LDaccel}) is
\begin{equation}
\gamma^2 \left( \langle V_i^2 \rangle + \frac{2 D}{\Delta t} \right)
=
\frac{(\langle V_i^2 \rangle)^3}{D^2}
+
\frac{2 (\langle V_i^2 \rangle)^2}{D \, \Delta t}. 
\label{LDaccelsec}
\end{equation} 
Using the MD values of $D$ and $\langle V_i^2 \rangle$
for K$^+$ which are given in Table~\ref{tableMDaverages}
and using MD timestep $\Delta t = 10^{-3}$ ps, we obtain
that the second moment~(\ref{LDaccelsec}) is equal to
$4.44 \times 10^5$ \AA$^2$ ps$^{-4}$. On the other
hand, $\langle U_i^2 \rangle$ estimated from all-atom MD
simulations and given in Table~\ref{tableMDaverages} is
$4.86 \times 10^3$ \AA$^2$ ps$^{-4}$ which is one hundred 
times smaller. The main reason for this discrepancy is
that Langevin dynamics postulates that the random force in 
equation~(\ref{BDVeq}) acting on the particle at time $t$ is not
correlated to the random force acting on the particle
at time $t+\Delta t$. However, this is not true for all-atom
MD simulations where random force terms at subsequent time 
steps are highly correlated. 

\newcommand{\picturesAB}[4]{
\centerline{
\hskip #4
\raise #3 \hbox{\raise 0.9mm \hbox{(a)}}
\hskip -8mm
\epsfig{file=#1,height=#3}
\raise #3 \hbox{\raise 0.9mm \hbox{(b)}}
\hskip -8mm
\epsfig{file=#2,height=#3}
}}
\newcommand{\picturesCD}[4]{
\centerline{
\hskip #4
\raise #3 \hbox{\raise 0.9mm \hbox{(c)}}
\hskip -8mm
\epsfig{file=#1,height=#3}
\raise #3 \hbox{\raise 0.9mm \hbox{(d)}}
\hskip -8mm
\epsfig{file=#2,height=#3}
}}

\begin{figure}[t]
\picturesAB{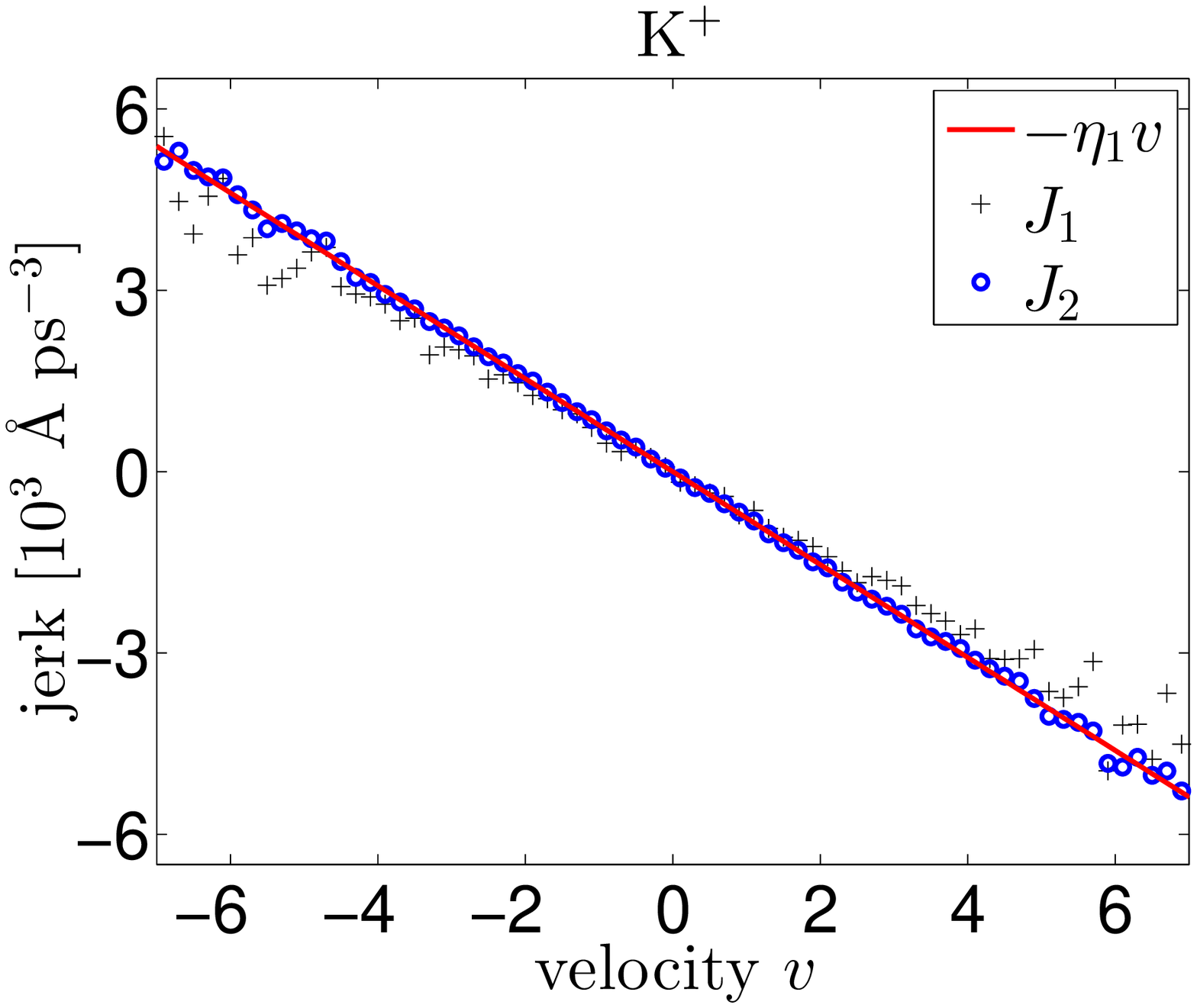}{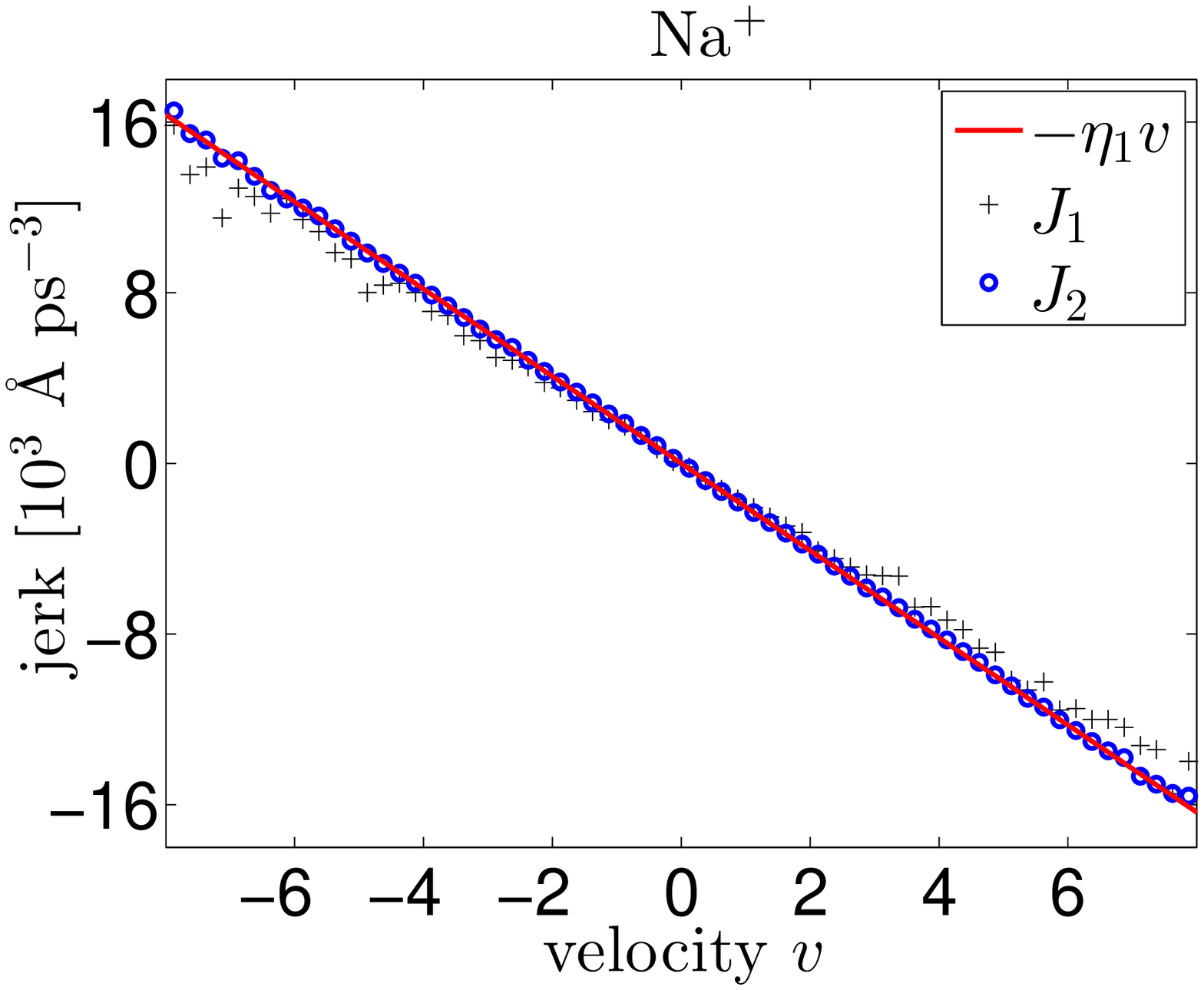}{5.3cm}{8mm}
\picturesCD{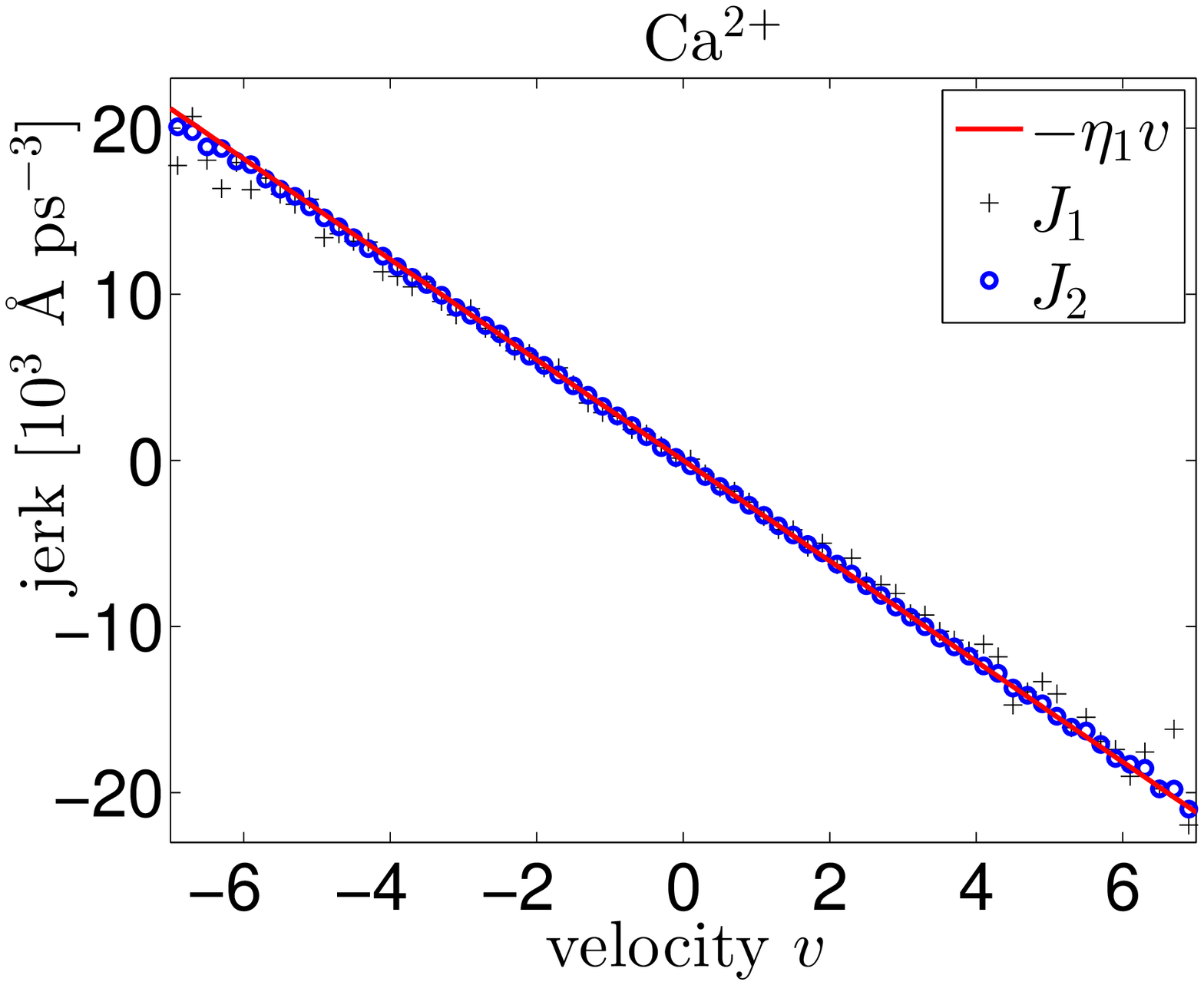}{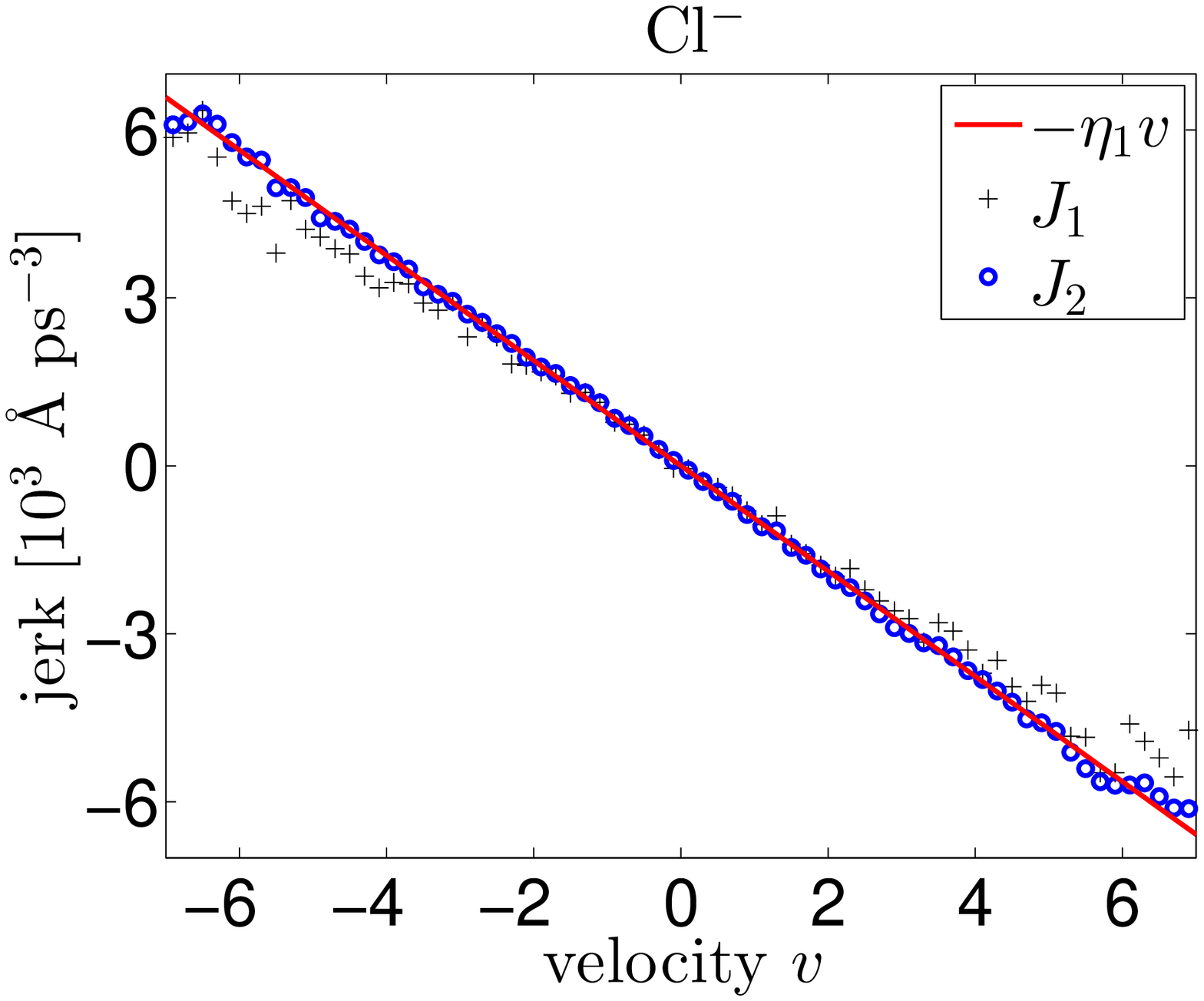}{5.3cm}{8mm}
\caption{{\it Comparison of the rate of change of 
acceleration (jerk) computed by all-atom MD simulations
and by the coarse-grained model~$(\ref{BDXeqAAA})$--$(\ref{BDZeqAAA})$.
MD results are visualized using functions $J_1(v)$ 
(black crosses) and $J_2(v)$ (blue circles) defined 
by equation~$(\ref{J1J2def})$. The result obtained by the
coarse-grained model is given by formula~$(\ref{Jcoarseeta1v})$ 
(red solid line). We consider} \hfill\break
(a) K$^+$ {\it ion}; 
(b) Na$^+$ {\it ion}; 
(c) Ca$^{2+}$ {\it ion and}
(d) Cl$^-$ {\it ion. Parameters are given
in Tables~$\ref{tableMDparam}$ and~$\ref{tableetavalues}$.}
}
\label{figure1}
\end{figure}

Since Langevin dynamics is not suitable for coupling MD and BD 
models, we need to introduce a stochastic model of ion behaviour which 
is more complicated than Langevin dynamics. The coarse-grained 
model~(\ref{BDXeqAAA})--(\ref{BDZeqAAA}) studied in this paper
is a relatively simple example of such a model. Its parametrization, 
discussed in Section~\ref{coarseion}, guarantees that the coarse-grained
model~(\ref{BDXeqAAA})--(\ref{BDZeqAAA}) well approximates 
all-atom MD simulations at steady state. They both have the
same value of diffusion constant $D$ and steady state values 
of $\langle V_i^2 \rangle$, $\langle U_i^2 \rangle$ and 
$\langle Z_i^2 \rangle$. Next, we show that the
coarse-grained model~(\ref{BDXeqAAA})--(\ref{BDZeqAAA}) also 
compares well with all MD simulations at shorter timescales.
We consider the rate of change of acceleration (jerk or the scaled 
derivative of force). We define the average jerk as a function 
of current velocity and acceleration of the ion:
\begin{equation}
J(v,u)
=
\lim_{\tau \to 0}
\frac{\langle U_i(t+\tau) - u 
\, | \, V_i(t)=v, U_i(t)=u \rangle}{\tau}.
\label{jdefinition}
\end{equation}
To estimate $J(v,u)$ from all-atom MD simulations, we calculate the rate 
of change of acceleration during each MD time step
\begin{equation}
J(v,u)
\approx
\frac{\langle U_i(t+\Delta t) - u 
\, | \, V_i(t)=v, U_i(t)=u \rangle}{\Delta t},
\label{jdefinitiondiscr}
\end{equation}
i.e. we run a long (nanosecond) MD simulation, calculate the values of
$(U_i(t+\Delta t)-U_i(t))/\Delta t$ during every time step
and record their average in two-variable array $J(v,u)$ 
indexed by binned values of $V_i(t)=v$ and $U_i(t)=u$.
Since the estimated $J(v,u)$ only weakly depends on $u$, we 
visualize our results in Figure~\ref{figure1}
using two functions of one variable, $v$, namely
\begin{equation}
J_1(v) = J(v,0),
\qquad
\quad
\mbox{and}
\qquad
\quad
J_2(v) = \int_{-\infty}^\infty J(v,u) \, p_u(u) \, \mbox{d} u,
\label{J1J2def}
\end{equation}
where $p_u(u)$ is the steady state distribution of $U_i$ estimated
from the same long time MD trajectory. As before, we use all three
dimensions to calculate the averages $J(v,u)$ and $p_u(u)$.
Function $J_1(v)$ (which gives jerk at the most likely value
of $U_i$) is plotted using crosses and function $J_2(v)$, the average
over $U_i$ variable, is plotted using circles in Figure~\ref{figure1}. 
In order to compare all-atom MD simulations with the coarse-grained 
model~(\ref{BDXeqAAA})--(\ref{BDZeqAAA}), we calculate the corresponding
jerk matrix $J(v,u)$ for the coarse-grained model. We denote
by $p(v,u,z)$ the stationary distribution of the stochastic 
process~(\ref{BDVeqAAA})--(\ref{BDZeqAAA}), i.e. 
$p(v,u,z) \, \mbox{d}v\, \mbox{d} u\, \mbox{d}z$ is the probability
that $V_i(t) \in [v,v+\mbox{d}v)$,
$U_i(t) \in [u,u+\mbox{d}u)$ and $Z_i(t) \in [z,z+\mbox{d}v)$.
Then the jerk matrix~(\ref{jdefinition})
of the coarse-grained model~(\ref{BDXeqAAA})--(\ref{BDZeqAAA}) is
\begin{equation*}
J(v,u)
=
\int_{-\infty}^\infty
\lim_{\tau \to 0}
\frac{\langle U_i(t+\tau) - u 
\, | \, V_i(t)=v, U_i(t)=u, Z_i(t)=z \rangle}{\tau} 
\, p(v,u,z) \, \mbox{d} z.
\end{equation*}
Using (\ref{BDUeqAAA}), we rewrite it as
\begin{equation}
J(v,u)
=
\int_{-\infty}^\infty
\left(
-
\eta_1 v
+
z \right)
\, p(v,u,z) \, \mbox{d} z.
\label{Jcoarse}
\end{equation}
The stationary distribution $p(v,u,z)$ of~(\ref{BDVeqAAA})--(\ref{BDZeqAAA})
is Gaussian with mean $[0,0,0]^T$ and stationary covariance matrix:
$$
\frac{1}{2 \eta_1 \eta_2 \eta_3}
\left(
\begin{matrix}
\eta_4^2 & 0 & 0 \\
0 & \eta_1 \eta_4^2 & 0 \\
0 & 0 & \eta_1 \eta_3 \eta_4^2
\end{matrix}
\right). 
$$
Consequently, equation~(\ref{Jcoarse}) implies
\begin{equation}
J(v,u) = - \eta_1 v. 
\label{Jcoarseeta1v}
\end{equation}
In Figure~\ref{figure1}, we plot~(\ref{Jcoarseeta1v}) using the red
solid line. The comparison with all atom MD results (circles and
squares) is excellent for all four ions considered in this paper.
In particular, we have shown that the coarse-grained
model~(\ref{BDXeqAAA})--(\ref{BDZeqAAA})
provides a good description of the rate of change of acceleration
(jerk) at the MD timescale. We make use of this property 
in Section~\ref{seccoupleMDBD} where 
we use the same time step ($\Delta t = 10^{-3}$ ps) for both
the coarse-grained model~(\ref{BDXeqAAA})--(\ref{BDZeqAAA})   
and all-atom MD simulations. The coarse-grained
model~(\ref{BDXeqAAA})--(\ref{BDZeqAAA}) can also be coupled with 
BD description~(\ref{BDSDE}), which uses much larger time steps, 
as we show in the next section.

\section{From the coarse-grained model (\ref{BDXeqAAA})--(\ref{BDZeqAAA}) 
to Brownian dynamics}
\label{seccoupleBDXVUZ}

\noindent
Let us consider the three-variable subsystem
(\ref{BDVeqAAA})--(\ref{BDZeqAAA}) of the coarse-grained model.
Denoting ${\mathbf y}_i = [V_i,U_i,Z_i]$,
equations~(\ref{BDVeqAAA})--(\ref{BDZeqAAA}) can be written 
in vector notation as follows
\begin{equation}
\mbox{d} {\mathbf y}_i = B \, {\mathbf y}_i \, \mbox{d}t
+ {\mathbf b} \, \mbox{d}W_i,
\label{vectorVUZsystem}
\end{equation} 
where matrix $B \in {\mathbb R}^{3 \times 3}$ and vector
${\mathbf b} \in {\mathbb R}^{3}$ are given as
\begin{equation}
B
=
\left(
\begin{matrix}
0 & 1 & 0 \\
-\eta_1 & 0 & 1 \\
0 & - \eta_3 & -\eta_2 
\end{matrix} 
\right)
\qquad
\mbox{and}
\qquad
{\mathbf b}
=
\left(
\begin{matrix}
0 \\
0 \\
\eta_4
\end{matrix} 
\right).
\label{defBb}
\end{equation} 
Let us denote the eigenvalues and eigenvectors of $B$ as 
$\lambda_j$ and 
${\boldsymbol \nu}_j = [\nu_{1j},\nu_{2j}, \nu_{3j}]$, 
$j=1,2,3$, respectively.
The eigenvalues of $B$ are the solutions of the characteristic 
polynomial 
$$
\lambda^3
+ \eta_2 \, \lambda^2 
+ (\eta_1 + \eta_3) \lambda 
+ \eta_1 \eta_2 = 0.
$$
Since $\eta_1$, $\eta_2$ and $\eta_3$ are positive parameters, we 
conclude that real parts of all three eigenvalues are negative 
and lie in interval $(-\eta_2,0).$ Using the values of
$\eta_j$, $j=1,2,3$, given in Table~\ref{tableetavalues}, 
we present the values of eigenvalues $\lambda_j$, $j=1,2,3,$ 
in Table~\ref{tableeigenvalues}.
\begin{table}
\centerline{
\begin{tabular}{|c||c|c|c|c|c|}
\hline 
\rule{0pt}{4.8mm}
& 
\raise 0.681mm \hbox{$\lambda_1$} 
& 
\raise 0.681mm \hbox{$\lambda_2$} 
& 
\raise 0.681mm \hbox{$\lambda_3$} 
& 
\raise 0.681mm \hbox{$t_1^*$} 
& 
\raise 0.681mm \hbox{$t_2^*$} 
\\
\rule{0pt}{4.8mm} \!\!\!\! 
\raise 0.681mm \hbox{ion} \!\!\!\!
& \!\!\!\raise 0.681mm \hbox{[ps$^{-1}$]}\!\!\! 
& \!\!\!\raise 0.681mm \hbox{[ps$^{-1}$]}\!\!\! 
& \!\!\!\raise 0.681mm \hbox{[ps$^{-1}$]}\!\!\! 
& \!\!\!\raise 0.681mm \hbox{[ps]}\!\!\! 
& \!\!\!\raise 0.681mm \hbox{[ps]}\!\!\! 
\\
\hline 
\hline 
\rule{0pt}{4.8mm}
\!\!\!\! K$^+$ \!\!\!\!& $\!\!\! -127.0$ \!\!\! 
&\!\!\! $-12.75 + 27.58 \, \mathbbm{i}$ \!\!\!
&\!\!\! $-12.75 - 27.58 \, \mathbbm{i}$ \!\!\!
&\!\!\! $3.08 \!\times\! 10^{-2}$ \!\!\! 
&\!\!\! $-9.39 \!\times\! 10^{-3}$ \!\!\!
\\
\hline 
\rule{0pt}{4.8mm}
\!\!\!\! Na$^+$ \!\!\!\! &\!\!\! $-140.1$ \!\!\!
&\!\!\! $-12.99 + 47.47 \, \mathbbm{i}$ \!\!\!
&\!\!\! $-12.99 - 47.47 \, \mathbbm{i}$ \!\!\!
& \!\!\! $6.15 \!\times\! 10^{-3}$ \!\!\!
&\!\!\! $-2.35 \!\times\! 10^{-2}$ \!\!\!\\
\hline 
\rule{0pt}{4.8mm}
\!\!\!\! Ca$^{2+}$ \!\!\!\! &\!\!\! $-163.1$ \!\!\!
&\!\!\! $-13.58 + 57.84 \, \mathbbm{i}$ \!\!\!
&\!\!\! $-13.58 - 57.84 \, \mathbbm{i}$ \!\!\!
&\!\!\! $1.47 \!\times\! 10^{-3}$ \!\!\!
&\!\!\! $-2.48 \!\times\! 10^{-2}$ \!\!\!\\
\hline 
\rule{0pt}{4.8mm}
\!\!\!\! Cl$^-$ \!\!\!\! &\!\!\! $-162.9$ \!\!\!
&\!\!\! $-13.41 + 30.25 \, \mathbbm{i}$ \!\!\!
&\!\!\! $-13.41 - 30.25 \, \mathbbm{i}$ \!\!\!
&\!\!\! $2.50 \!\times\! 10^{-2}$ \!\!\!
&\!\!\! $-1.13 \!\times\! 10^{-2}$ \!\!\!\\
\hline
\end{tabular}
}
\caption{{\it Eigenvalues $\lambda_j$, $j=1,2,3$, of matrix
$B$ defined by~$(\ref{defBb})$ and time shifts
$t_1^*$ and $t_2^*.$ \hfill\break Symbol $\mathbbm{i}$ denotes 
the imaginary unit.} \label{tableeigenvalues}}
\end{table}%
The eigenvalues $\lambda_j$, $j=1,2,3$, are distinct. The general
solution of the SDE system (\ref{vectorVUZsystem}) can be written
as follows~\cite{Mao:2007:SDE}
\begin{equation}
{\mathbf y}_i(t)
=
\Phi(t) \, {\mathbf c}
+
\Phi(t) \int_0^t \Phi^{-1}(s) \, {\mathbf b} \, \mbox{d}W_i,
\label{gensolutionVUZsystem}
\end{equation} 
where ${\mathbf c} \in {\mathbb R}^{3}$ is a constant vector 
determined by initial conditions and matrix
$\Phi(t) \in {\mathbb R}^{3 \times 3}$
is given as
$
\Phi(t) 
= 
[\exp(\lambda_1 t) {\boldsymbol \nu}_1 \; | \;
 \exp(\lambda_2 t) {\boldsymbol \nu}_2 \; | \;
 \exp(\lambda_3 t) {\boldsymbol \nu}_3],
$
i.e. each column is a solution of the ODE system
$\mbox{d} {\mathbf y}_i = B \, {\mathbf y}_i \, \mbox{d}t$.
Considering deterministic initial conditions, 
equation~(\ref{gensolutionVUZsystem}) implies that the
process is Gaussian at any time $t>0$. Equations for means, 
variances and covariances then uniquely determine the distribution
of ${\mathbf y}_i(t)$ for $t > 0$. Equations for means can be written
in the vector form as 
$\mbox{d} \langle {\mathbf y}_i \rangle = B \, 
\langle {\mathbf y}_i \rangle \, \mbox{d}t.$
Equations for variances and covariances are given in 
Section~\ref{coarseion} as equations~(\ref{V2eq})--(\ref{VZeq}),
(\ref{UZeq}), (\ref{X2eq})--(\ref{XZeq}) and (\ref{Z2eq}).

\begin{figure}[t]
\vskip 2.5mm
\leftline{
\hskip -1.6mm
\epsfig{file=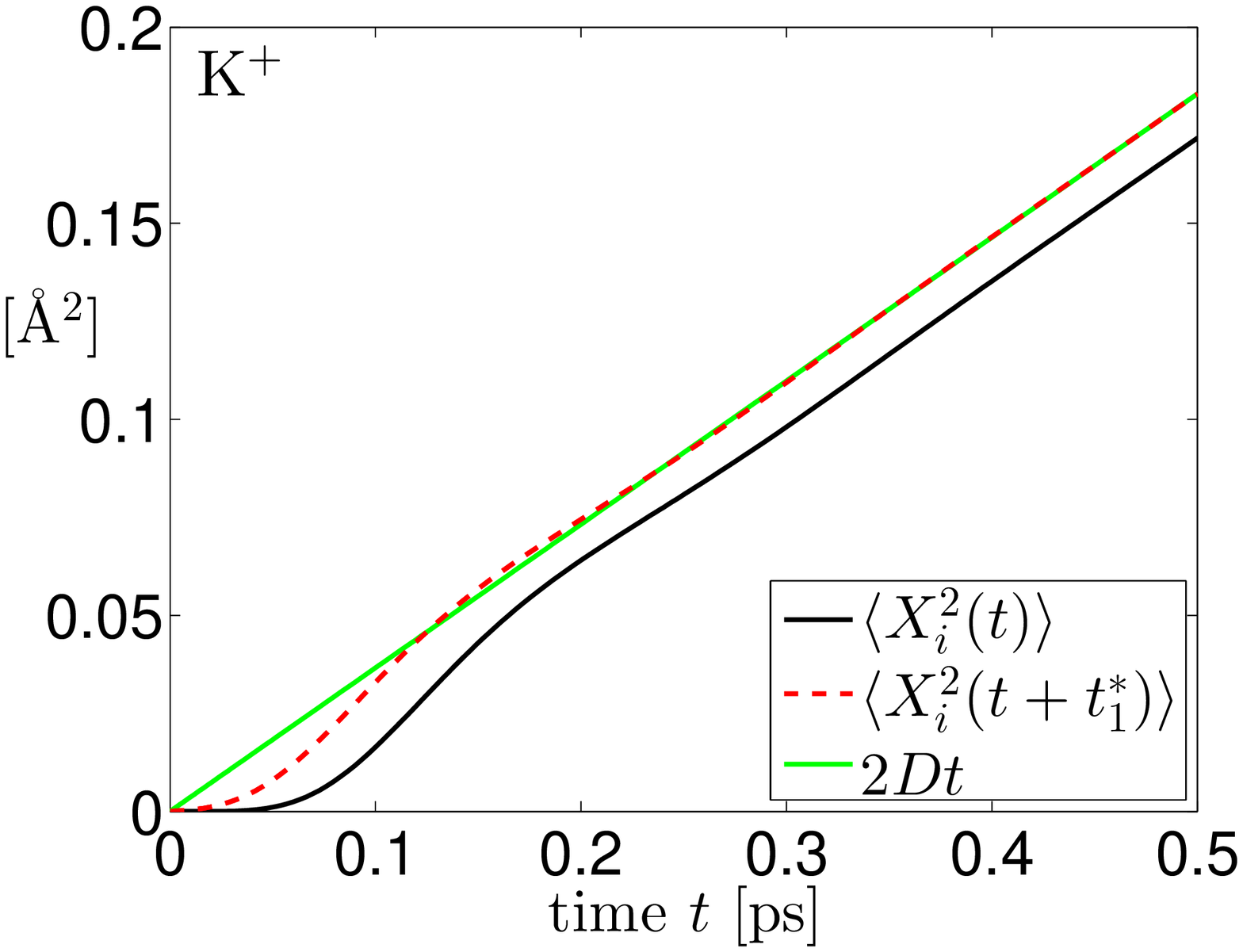,height=5.25cm}
\hskip -2mm
\epsfig{file=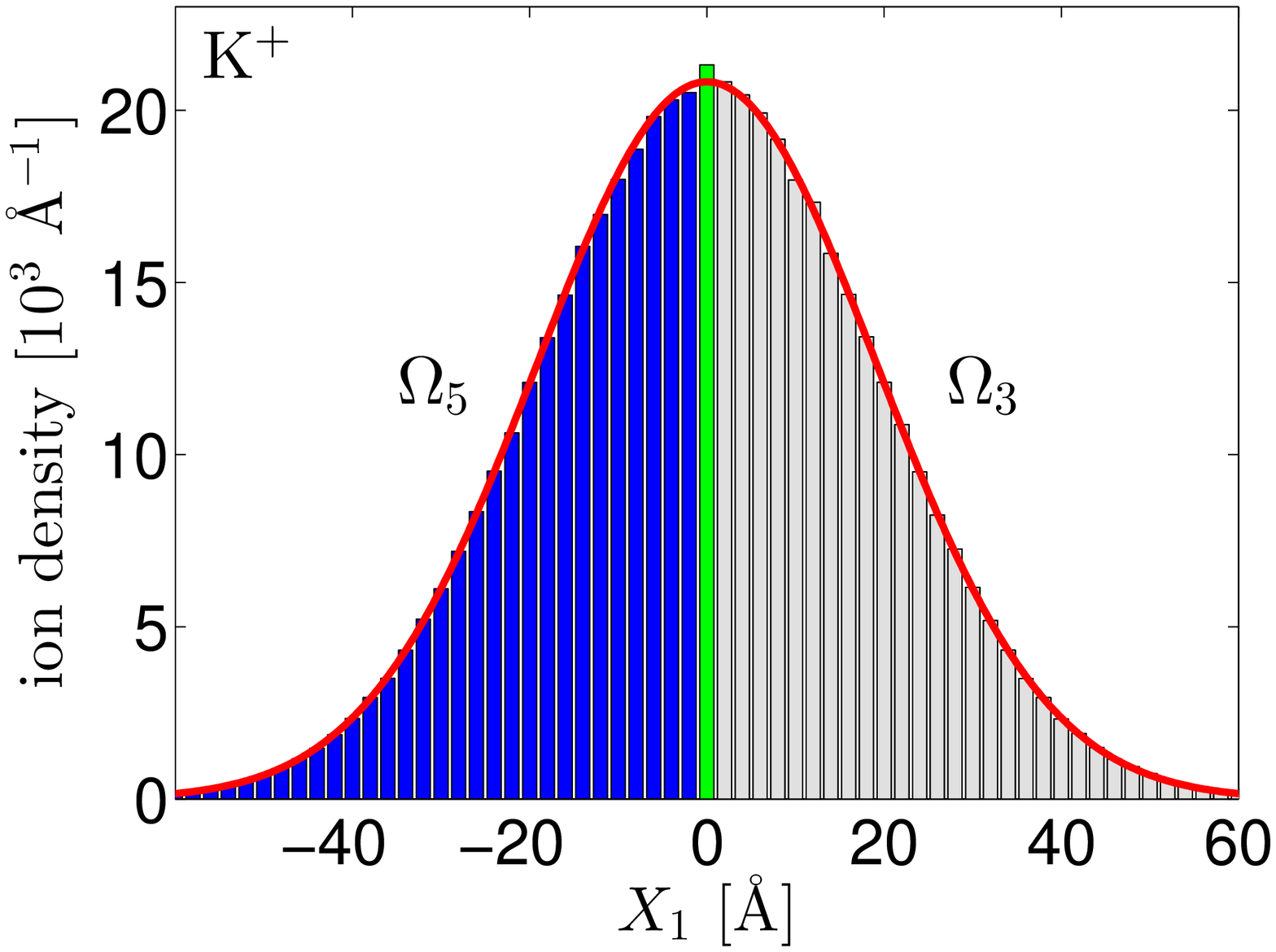,height=5.35cm}
}
\vskip -5.75cm
\leftline{(a) \hskip 6.3cm (b)}
\vskip 5.3cm
\caption{(a) {\it Comparison of the coarse-grained
model~$(\ref{BDXeqAAA})$--$(\ref{BDZeqAAA})$ and
BD description~$(\ref{BDSDE})$ for {\rm K}$^+$ ion.
The mean square displacement computed by solving 
10 ODEs~$(\ref{V2eq})$--$(\ref{VZeq})$, $(\ref{UZeq})$, 
$(\ref{X2eq})$--$(\ref{XZeq})$ and $(\ref{Z2eq})$ with zero
initial conditions (black solid line). The same
curve shifted by the value of $t_1^*$ is plotted
as a red dashed line.}
(b) {\it Test of accuracy of the multiscale approach
in~$\Omega_3 \cup \Omega_4 \cup \Omega_5$ for {\rm K}$^+$ ion. 
Histogram obtained by simulating $10^6$ ions over time
$10^3\,${\rm ps} is compared with 
the analytical result~$(\ref{exsol106})$ (red solid line).
Grey bars show the ion density in $\Omega_3$, the green bar
shows the ion density in $\Omega_4$ and blue bars show the
ion density in $\Omega_5$.
Parameters are given in Tables~$\ref{tableMDparam}$
and~$\ref{tableetavalues}$.}
}
\label{figure2}
\end{figure}

There are two important conclusions of the above analysis.
First of all, eigenvalues $\lambda_j$, $j=1,2,3$,
given in Table~\ref{tableeigenvalues} satisfy
$$
\lambda_1 < \mbox{Re} \, \lambda_2 = \, \mbox{Re} \lambda_3 < 0, 
$$ 
where Re denotes the real part of a complex number. There is 
a spectral gap between the first eigenvalue and the complex 
conjugate pair of eigenvalues.
If we used this spectral gap, we could reduce the system to two
evolution equations for times $t \gg 1/|\lambda_1|$. However, there
is no spectral gap to reduce the system to 
Langevin dynamics~(\ref{BDXeq})--(\ref{BDVeq}). In particular,
we again confirm our conclusion that a coarse-grained approximation 
of ion behaviour is not given in terms of Langevin dynamics.
Our second conclusion is that on a picosecond time scale, 
we can assume stationarity in~(\ref{vectorVUZsystem}) 
to get
\begin{eqnarray}
\mbox{d}X_i & = & 
\frac{\eta_4}{\eta_1 \eta_2} 
\; \mbox{d}W_i, 
\quad
i=1,2,3.
\label{BDXeqred}
\end{eqnarray}
Using (\ref{eta1estimate}), (\ref{eta2estimate}) and (\ref{eta4estimate}),
we have
$$
\frac{\eta_4}{\eta_1 \eta_2} = \sqrt{2 D}.
$$
Consequently, equation~(\ref{BDXeqred}) is equivalent to
BD description~(\ref{BDSDE}). The convergence of
(\ref{BDXeqAAA})--(\ref{BDZeqAAA})
to the BD model is illustrated in Figure~\ref{figure2}(a). We solve
the system of 10 ODEs for variances and covariances given as
equations~(\ref{V2eq})--(\ref{VZeq}),
(\ref{UZeq}), (\ref{X2eq})--(\ref{XZeq}) and (\ref{Z2eq}).
We consider (deterministic) zero initial conditions,
i.e. $X_i(0)=V_i(0)=U_i(0)=Z_i(0)=0$. All moments are then
initially equal to zero. We plot the mean square displacement 
$\langle X_i^2 \rangle$ as a function of time.
We compare it with the mean square displacement of BD
model~(\ref{BDSDE}) which is given as $2Dt.$ We observe
that there is an approximately constant shift, denoted
$t_1^*$, between both solutions for times $t > 0.2$ ps. 
We illustrate this further by plotting $\langle X_i^2(t+t_1^*) \rangle$
in Figure~\ref{figure2}(a). The values of shift $t_1^*$ for different 
ions estimated by solving the ODEs for second moments with zero initial
conditions are given in Table~\ref{tableeigenvalues}.

Next, we show how the BD model~(\ref{BDSDE}) and the coarse-grained
model~(\ref{BDXeqAAA})--(\ref{BDZeqAAA}) can be used in different
parts of the computational domain. This coupling will form one
component of multiscale methodology developed in Section~\ref{seccoupleMDBD}.
BD algorithms based on equation~(\ref{BDSDE}) have been 
implemented in a number of methods designed for spatio-temporal 
modelling of intracellular processes, including 
Smoldyn~\cite{Andrews:2004:SSC}, MCell~\cite{Stiles:2001:MCM} and 
Green's-function reaction dynamics~\cite{vanZon:2005:GFR}. 
Smoldyn discretizes~(\ref{BDSDE}) using a fixed 
BD time step $\Delta T$, i.e. it computes the time evolution of the 
position ${\mathbf X} \equiv {\mathbf X}(t)$ of each molecule by 
\begin{equation}
X_i(t+\Delta t) = X_i(t) + \sqrt{2D \Delta T} \, \xi_{i},
\qquad
i=1,2,3,
\label{discBD}
\end{equation}
where $[\xi_{1},\xi_{2},\xi_{3}]$ is a vector of normally distributed 
random numbers with zero mean and unit variance. We use
discretization~(\ref{discBD}) of BD model~(\ref{BDSDE}) in this paper.
BD time step $\Delta T$ has to be chosen much larger than the
MD time step $\Delta t$. We use $\Delta T = 0.5$ ps, but any
larger time step would also work well. We could also use a
variable time step, as implemented in the Green's Function 
Reaction Dynamics \cite{vanZon:2005:GFR}. 

In Section~\ref{seccoupleMDBD}, we consider all-atom MD simulations
in domain $\Omega \subset {\mathbb R}^3$. Our main goal is to design 
a multiscale approach which can compute spatio-temporal 
statistics with the MD-level of detail in relatively small subdomain 
$\Omega_{1} \subset \Omega$ by using BD model~(\ref{discBD}) in 
the most of the rest of the computational domain. This is achieved by 
decomposing domain $\Omega$ into five subdomains 
$\Omega_{j}$, $j=1,2,3,4,5$ (see equation~(\ref{om15def})
and discussion in Section~\ref{seccoupleMDBD}).
We use MD in $\Omega_1$, the coarse-grained
model~(\ref{BDXeqAAA})--(\ref{BDZeqAAA})
in $\Omega_3$ and the BD model~(\ref{discBD}) in $\Omega_5$. 
The remaining two subdomains, $\Omega_2$ and $\Omega_4$, are two
overlap (hand-shaking) regions where two different simulation
approaches can be used at the same time~\cite{Erban:2014:MDB,Franz:2013:MRA}. 
In the rest of this section, we focus on simulations in 
region $\Omega_3 \cup \Omega_4 \cup \Omega_5$
which concerns coupling the coarse-grained 
model~(\ref{BDXeqAAA})--(\ref{BDZeqAAA})  
with the BD model~(\ref{discBD}). We use the
coarse-grained model in $\Omega_{3} \cup \Omega_{4}$
and the BD model~(\ref{discBD}) in $\Omega_{4} \cup \Omega_{5}$.
In particular, we use both models in the overlap region 
$\Omega_{4}$. Each particle which is initially in $\Omega_{3}$
is simulated according to~(\ref{BDXeqAAA})--(\ref{BDZeqAAA})  
(discretized using time step $\Delta t$)
until it enters $\Omega_{5}$. Then we use~(\ref{discBD})
to evolve the position of a particle (over BD time steps of length
$\Delta T$) until it again enters $\Omega_{3}$ when we switch
the description back from the BD model to the coarse-grained model.
In order to do this, we have to initialize variables 
$V_i$, $U_i$ and $Z_i$, $i=1,2,3$. We use deterministic
initial conditions, $V_i(0)=U_i(0)=Z_i(0)=0$, disccused above. 

In Figure~\ref{figure2}(b), we present an illustrative simulation
where $\Omega_3 \cup \Omega_4 \cup \Omega_5 = {\mathbb R}^3$
for simplicity. We use $\Omega_3 = (h,\infty) \times {\mathbb R}^2$, 
$\Omega_4 = [-h,h] \times {\mathbb R}^2$ and 
$\Omega_5 = (-\infty,-h) \times {\mathbb R}^2$,
where $h = 1\,$\AA. We report averages over $10^6$ simulations 
of ions, half of them are initiated at ${\mathbf X}(0)=[h,0,0]$, 
i.e. they initially follow the coarse-grained
model~(\ref{BDXeqAAA})--(\ref{BDZeqAAA}) with zero
initial condition for other variables
($V_i(0)=U_i(0)=Z_i(0)=0$). The second half of ions
are initiated at ${\mathbf X}(0)=[-h,0,0]$, i.e. they
initially follow BD description~(\ref{discBD}).
We plot the (marginal) distribution of ions along the first 
coordinate ($X_1$) at time $10^3\,$ps in Figure~\ref{figure2}(b).
The computed histogram is plotted using bins of length
$2\,$\AA, i.e. the overlap region $\Omega_4$ is equal to one bin 
(visualized as a green bar). Grey (resp. blue) bars show the density 
of ions in $\Omega_3$ (resp. $\Omega_5$). We compare our results 
with the analytical distribution computed for BD description~(\ref{BDSDE})
at time $t=10^3\,$ps given by
\begin{equation}
\varrho(x_1)
=
\frac{10^6}{4 \sqrt{\pi D t}}
\left(
\exp \left[
- \frac{(x_1 - h)^2}{4 D t}
\right]
+
\exp \left[
- \frac{(x_1 + h)^2}{4 D t}
\right]
\right).
\label{exsol106}
\end{equation}
The computed histogram compares well with~(\ref{exsol106}), although we
can observe a small error: the green bar is slightly taller than
the corresponding value of~(\ref{exsol106}). If we wanted
to further improve the accuracy, we could take into account that there 
is time shift $t_1^*$, discussed above, introduced to the multiscale
approach by using the deterministic initial conditions,
$V_i(0)=U_i(0)=Z_i(0)=0$, 
for ions entering domain $\Omega_3$. Another possibility is to sample 
the initial condition for $V_i,$ $U_i$ and $Z_i$ from a suitable 
distribution. If we use the stationary distribution of subsystem
(\ref{BDVeqAAA})--(\ref{BDZeqAAA}), then $\langle V_i^2 \rangle$ does 
not evolve and is equal to
$$
\langle V_i^2 \rangle = \frac{\eta_4^2}{2 \eta_1 \eta_2 \eta_3}.
$$
Substituting this constant for $\langle V_i^2 \rangle$ into~(\ref{XVeq}), 
the system of 10 ODEs for second moments
of~(\ref{BDXeqAAA})--(\ref{BDZeqAAA})
simplifies to 4 ODEs~(\ref{X2eq})--(\ref{XZeq}). Solving 
system~(\ref{X2eq})--(\ref{XZeq})
with zero initial conditions (assuming $X_i(0)=0$), we can 
again compute the mean square displacement. As in 
Figure~\ref{figure2}(a), it can be shifted in time to better
match with the BD result, $2Dt$. We denote this time shift as $t_2^*$. 
Its values are given in Table~\ref{tableeigenvalues}. We observe 
that $t_2^*$ is negative and $t_1^*$ is positive for all four ions 
considered in Table~\ref{tableeigenvalues}. Both time shifts $t_1^*$ 
and $t_2^*$ (together with optimizing size $h$ of the overlap region) 
could be used to further improve the accuracy of multiscale simulations 
in~$\Omega_3 \cup \Omega_4 \cup \Omega_5$~\cite{Erban:2014:MDB}.
However, our main goal is to introduce a multiscale approach 
which can use all-atom MD simulations in $\Omega_1$. Since MD
simulations are computationally intensive, we will only consider
100 realizations of the multiscale method in Section~\ref{seccoupleMDBD}. 
In particular, the Monte Carlo error will be larger than the error observed 
in Figure~\ref{figure2}(b). Thus, we can use the above approach
in~$\Omega_3 \cup \Omega_4 \cup \Omega_5$ without introducing 
observable errors in the multiscale method developed in the
next section.

\section{Coupling all-atom MD and BD}
\label{seccoupleMDBD}

\begin{figure}[t]
\leftline{
\epsfig{file=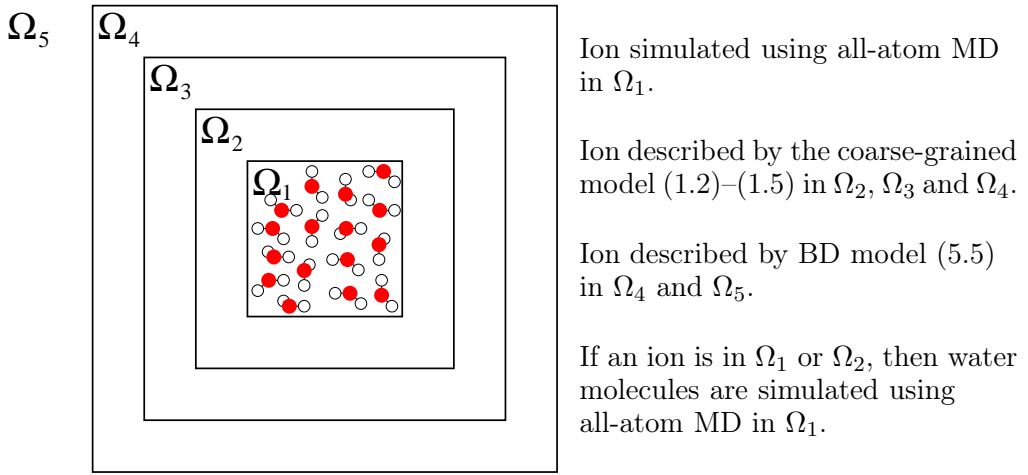,height=6.2cm}
}
\vskip -5.9cm 
\hskip 7.7cm Ion simulated using all-atom MD
\vskip 0.01cm
\hskip 7.7cm in $\Omega_1$.
\vskip 5 mm
\hskip 7.7cm Ion described by the coarse-grained 
\vskip 0.01cm
\hskip 7.7cm model~(\ref{BDXeqAAA})--(\ref{BDZeqAAA})
in $\Omega_2$, $\Omega_3$ and $\Omega_4$.
\vskip 5 mm
\hskip 7.7cm Ion described by BD model~(\ref{discBD})
\vskip 0.01cm
\hskip 7.7cm in $\Omega_4$ and $\Omega_5$. 
\vskip 5 mm
\hskip 7.7cm If an ion is in $\Omega_1$ or $\Omega_2$, then water
\vskip 0.01cm
\hskip 7.7cm molecules are simulated using 
\vskip 0.01cm
\hskip 7.7cm all-atom MD in $\Omega_1$.
\vskip 4 mm
\caption{{\it Schematic of multiscale set up. Note that the schematic 
is drawn in two spatial dimensions to enable better visualization, 
but all models are formulated and simulated in three spatial dimensions.}}
\label{figure3}
\end{figure}

Let us consider all-atom MD in domain $\Omega \subset {\mathbb R}^3$ 
which is so large that direct MD simulations would be too computationally 
expensive. Let us assume that a modeller only needs to consider
the MD-level of detail in a relatively small subdomain 
$\Omega_1 \subset \Omega$ while, in the rest of the computational
domain, ions are transported by diffusion and BD 
description~(\ref{BDSDE}) is applicable. For example, 
domain $\Omega_1$ could include binding sites for ions or (parts of) 
ion channels. In this paper, we do not focus on a specific application. 
Our goal is to show that the coarse-grained model~(\ref{BDXeqAAA})--(\ref{BDZeqAAA}) is
an intermediate model between all-atom MD and BD which enables 
the use of both methods during the same dynamic simulation. To achieve this,
we decompose domain $\Omega$ into five subdomains, denoted
$\Omega_j$, $j=1,2,3,4,5,$ see Figure~\ref{figure3}. These sets are
considered pairwise disjoint (i.e. $\Omega_i \cap \Omega_j = \emptyset$ 
for $i \ne j$) and 
\begin{equation}
\Omega = \Omega_1 \cup \Omega_2 \cup \Omega_3 \cup \Omega_4 \cup \Omega_5.
\label{om15def}
\end{equation}
In our illustrative simulations, we consider the behaviour of one ion.
If the ion is in subdomain $\Omega_1$, then we use all-atom MD
simulations as described in Section~\ref{MDsim}. In particular,
the force between the ion and a water molecule is obtained by 
differentiating potential~(\ref{pairpotMD}), provided that the
distance between the ion and the water molecule is less than
the cutoff distance ($L/2$). Let us denote the
force exerted by the ion on the water molecule by 
${\mathbf F}_{iw} (r_{i0},r_{i1},r_{i2})$,
where $r_{i0}$ (resp., $r_{i1}$ and $r_{i2}$) is the distance 
between the ion and the oxygen site (resp., the first 
and second hydrogen sites) on the water molecule.
We use periodic boundary conditions for water molecules in $\Omega_1$. 

Whenever the ion leaves $\Omega_1$, it enters $\Omega_2$ where we simulate 
its behaviour using the coarse-grained 
model~(\ref{BDXeqAAA})--(\ref{BDZeqAAA}). We still simulate water molecules 
in $\Omega_1$ and we allow them to experience additional forces exerted 
by the ion which is present in $\Omega_2$. These forces have the same 
functional form, ${\mathbf F}_{iw}$, as in MD, but they have modified 
arguments as follows
\begin{equation}
{\mathbf F}_{iw} 
\big(
r_{i0} + \omega \, \mbox{dist} ({\mathbf X}, \Omega_1),
r_{i1} + \omega \, \mbox{dist} ({\mathbf X}, \Omega_1),
r_{i2} + \omega \, \mbox{dist} ({\mathbf X}, \Omega_1)
\big),
\label{iwforce}
\end{equation}
where $\omega \ge 0$ is a parameter and 
$\mbox{dist} ({\mathbf X}, \Omega_1)$ is the (closest)
distance between the ion at position ${\mathbf X}$ and 
subdomain $\Omega_1$. 
If the ion is in region $\Omega_3 \cup \Omega_4 \cup \Omega_5$, 
then water molecules in $\Omega_1$ are no longer simulated.
We use the coarse-grained model~(\ref{BDXeqAAA})--(\ref{BDZeqAAA})
to simulate the ion behaviour in $\Omega_3$ and the
BD model~(\ref{discBD}) in $\Omega_5$. Overlap region $\Omega_4$
is used to couple these simulation methods as explained in 
Section~\ref{seccoupleBDXVUZ}.

In Section~\ref{seccoupleBDXVUZ}, we have already presented
illustrative simulations to validate the multiscale modelling 
strategy chosen in region $\Omega_3 \cup \Omega_4 \cup \Omega_5$. 
Next, we focus on testing and explaining the multiscale approach 
chosen to couple region $\Omega_1$ with $\Omega_2$. The key
idea is given by force term~(\ref{iwforce}) which is used
for MD simulations of water molecules in $\Omega_1$
when an ion is in $\Omega_2$. This force term has two important
properties:

\medskip

{

\leftskip 15mm

\parindent -5.4mm

(i) If an ion is on the boundary of $\Omega_1$, i.e. 
${\mathbf X} \in \partial \Omega_1$, then 
$\mbox{dist} ({\mathbf X}, \Omega_1) = 0$ and 
force~(\ref{iwforce}) is equal to force term
${\mathbf F}_{iw} (r_{i0},r_{i1},r_{i2})$ used 
in $\Omega_1$.

\medskip

\parindent -6.6mm

(ii) If $\omega \, \mbox{dist} ({\mathbf X}, \Omega_1) \ge L/2$, 
then force~(\ref{iwforce}) is equal to zero.

\leftskip 0mm

}

\medskip

\noindent
Property (i) implies that formula~(\ref{iwforce}) continuously extends 
the force term used in MD. In particular, water molecules do not 
experience abrupt changes of forces when the ion crosses boundary
$\partial \Omega_1.$ Property (ii) is a consequence of the cutoff 
distance used (together with the reaction field
correction~\cite{Koneshan:1998:SSD}) 
to treat long-range interactions. In our illustrative simulations, 
we use 
\begin{equation}
\Omega_1 = \left[ - \frac{L}{2}, \frac{L}{2} \right]^3
\qquad \quad
\mbox{and}
\qquad \quad
\Omega_2 = \left[ - \frac{L}{2} - \frac{L}{2 \omega}, 
\frac{L}{2} + \frac{L}{2 \omega} \right]^3 \setminus \Omega_1.
\label{defom12}
\end{equation}
Property (ii) implies that extra force~(\ref{iwforce}) is equal to
zero on boundary $\partial \Omega_2 \setminus \partial \Omega_1$
which is the boundary between regions $\Omega_2$ and $\Omega_3$.
This is consistent with the assumption that ions in region 
$\Omega_3 \cup \Omega_4 \cup \Omega_5$ do not interact with 
water molecules in region $\Omega_1$. 

If an ion is in $\Omega_1$, we use all-atom MD as formulated in
Section~\ref{MDsim}. Periodic boundary conditions are implemented
in MD simulations. Water molecules are subject to forces exerted 
by the ion at its real position ${\mathbf X}$ in $\Omega_1$, but 
also by its copies at periodic locations 
${\mathbf X} + (i L,j L,k L)$ where $i,j,k \in {\mathbb Z}.$
When the ion moves to $\Omega_2$, one of its copies is in $\Omega_1$. 
Force term~(\ref{iwforce}) is designed in such a way, that the 
strength of interaction decreases (for every copy of the ion) 
with the distance, $\mbox{dist} ({\mathbf X}, \Omega_1)$, 
between the real position of the ion and $\Omega_1$. In particular, 
force term~(\ref{iwforce}) ensures that there are continuous
changes of all forces when the ion moves between regions
$\Omega_1$, $\Omega_2$ and $\Omega_3.$

\begin{figure}[t]
\vskip 2.5mm
\leftline{
\hskip -2mm
\epsfig{file=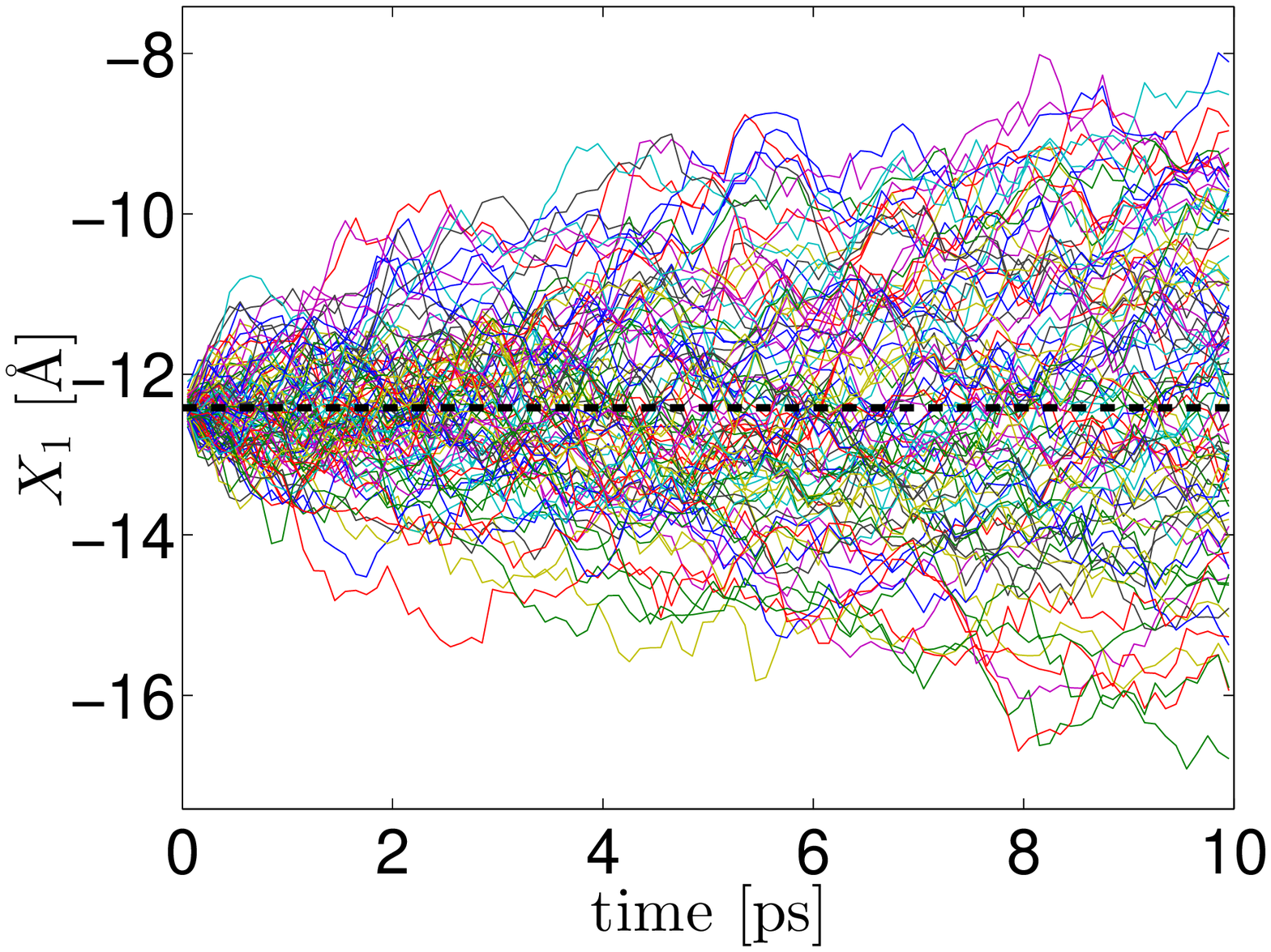,height=5.25cm}
\hskip -2mm
\epsfig{file=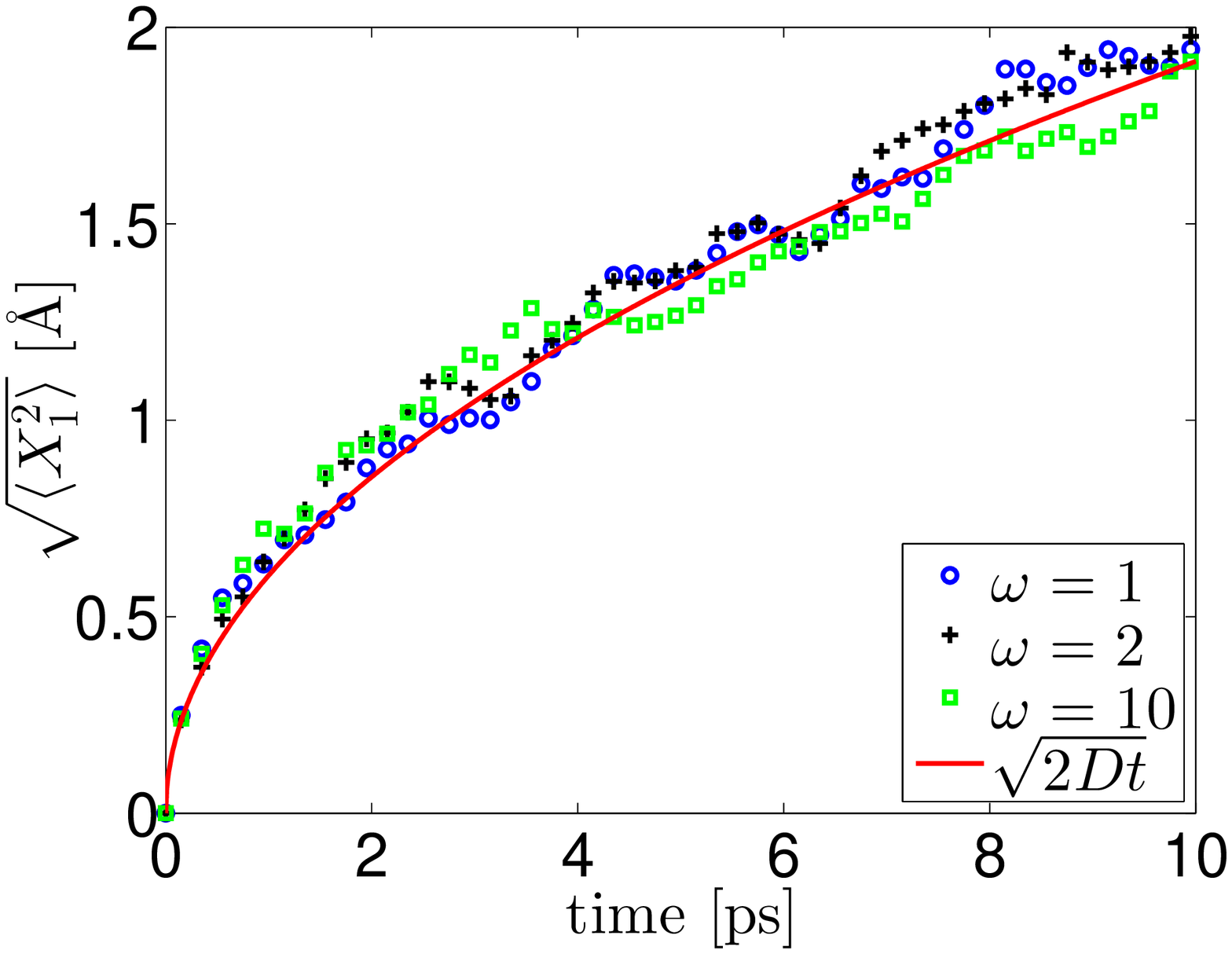,height=5.35cm}
}
\vskip -5.75cm
\leftline{(a) \hskip 6.3cm (b)}
\vskip 5.3cm
\caption{(a) {\it One hundred realizations of a multiscale simulation 
of K$^+$ ion initiated at $[-L/2,0,0]$. We plot $X_1$ coordinate as 
a function of time. Ion is described by all-atom MD for 
$X_1 \ge -L/2$ and by the coarse-grained
model~$(\ref{BDXeqAAA})$--$(\ref{BDZeqAAA})$ for $X_1 < -L/2$.
The boundary between $\Omega_1$ and $\Omega_2$ is visualized using
the black dashed line. We use $\omega=1$ in~$(\ref{iwforce})$.}
(b) {\it The mean square displacement in the first coordinate
of K$^+$ ion simulated in $\Omega_1 \cup \Omega_2$ and computed 
as the average of $100$ realizations for $\omega=1$ (blue circles),
$\omega=2$ (black crosses) and $\omega=10$ (green squares).}
}
\label{figure4}
\end{figure}%

In Figure~\ref{figure4}, we present results of simulations of K$^+$
ion in region $\Omega_1 \cup \Omega_2$. We consider 100 realizations
of a multiscale simulation with one ion. Its initial position is 
${\mathbf X}(0) = [-L/2,0,0]$ which lies on boundary $\partial \Omega_1$.
We simulate each realization for time 10$\,$ps which is short
enough that all trajectories stay inside the ball of radius $L/2$
centred at ${\mathbf X}(0)$. Then $X_1$-coordinate of the trajectory
determines whether the ion is in $\Omega_1$ or $\Omega_2$.
If $X_1(t) \ge -L/2$, then the ion is in $\Omega_1$ and it is simulated
using all-atom MD. If $X_2(t) < -L/2$, then the ion is in $\Omega_2$
and evolves according to the coarse-grained
model~(\ref{BDXeqAAA})--(\ref{BDZeqAAA}). In Figure~\ref{figure4}(a),
we use~(\ref{iwforce}) with $\omega=1$ and plot $X_1$ coordinates of all 
100 realizations. We observe that the computed trajectories spread on 
both sides of boundary $\partial \Omega_1$ (dashed line) without any
significant bias. The mean square displacement is presented in
Figure~\ref{figure4}(b) for three different values of $\omega$. 
The results compare well with $(2 D t)^{1/2}$ which is the mean 
square displacement of one coordinate of the diffusion process.

We conclude with illustrative simulations which are coupling all-atom 
MD with BD. We use domain $\Omega \in {\mathbb R}^3$ decomposed into
five regions as in equation~(\ref{om15def}), where 
$\Omega_1$ and $\Omega_2$ are given by (\ref{defom12}), and
\begin{eqnarray}
\Omega_3 &=& \left[ - \frac{L}{2} - \frac{L}{2\omega} - h_1, 
\frac{L}{2} + \frac{L}{2\omega} + h_1 \right]^3 
\setminus 
(\Omega_1 \cup \Omega_2),
\label{defom3}
\\
\Omega_4 &=& \left[ - \frac{L}{2} - \frac{L}{2\omega} - h_1 - h_2, 
\frac{L}{2} + \frac{L}{2 \omega}  + h_1 + h_2 \right]^3 
\setminus 
(\Omega_1 \cup \Omega_2 \cup \Omega_3),
\qquad
\label{defom4}
\\
\Omega_5 &=& {\mathbb R}^3 \setminus 
(\Omega_1 \cup \Omega_2 \cup \Omega_3 \cup \Omega_4),
\label{defom5}
\end{eqnarray}
where $\omega=10$, $h_1 = L/20$ and $h_2 = L/10$. Then the BD domain
is $\Omega_5 = {\mathbb R}^3 \setminus [ -7 L/10, 7 L/10]^3$.
We place an ion at the origin (centre of MD domain $\Omega_1$), 
i.e. ${\mathbf X}(0) = [0,0,0],$ and we simulate each trajectory
until it reaches the distance $4 L = 99.32\,$\AA{} from the origin.
Let ${\cal T}(r)$ be the time when a trajectory first reaches 
distance $r$ from the origin.
\begin{figure}[t]
\vskip 2.5mm
\leftline{
\hskip -2mm
\epsfig{file=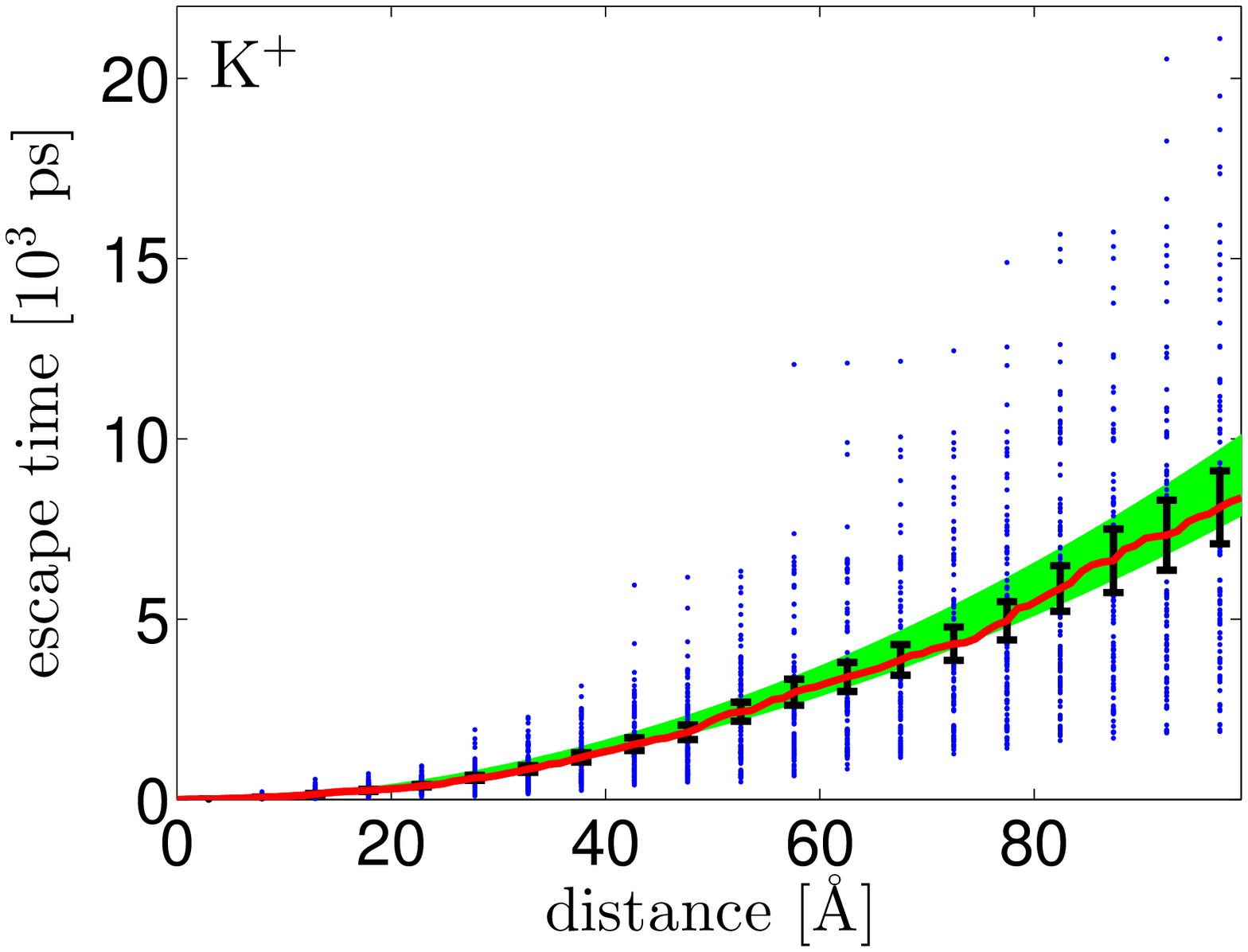,height=5.25cm}
\hskip -2mm
\epsfig{file=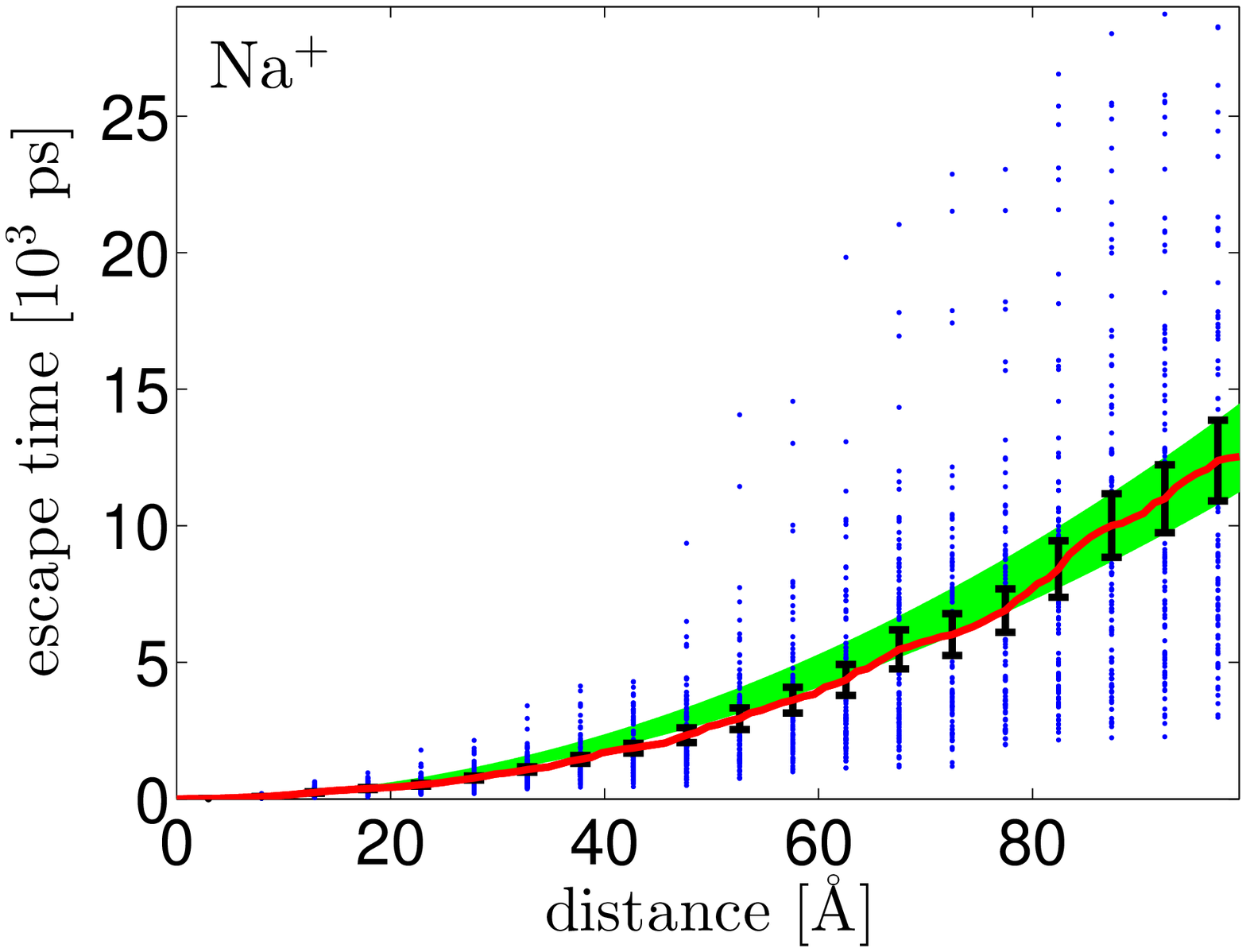,height=5.35cm}
}
\vskip -5.75cm
\leftline{(a) \hskip 6.3cm (b)}
\vskip 5.3cm
\caption{ {\it
Escape time ${\cal T}(r)$ to reach distance $r$ from the origin
computed by the multiscale method. We consider}
(a) K$^+$ {\it ion; and}
(b) Na$^+$ {\it ion.} {\it We plot escape times for individual
realizations (blue points), the mean escape time estimated
from $100$ realizations (red solid line) and the theoretical
$95$\% confidence interval~$(\ref{theorconfinterval})$
(green area). We use $\omega=10$ in~$(\ref{iwforce})$.}
}
\label{figure5}
\end{figure}%
In Figure~\ref{figure5}, we plot escape time ${\cal T}(r)$ as a function
of distance $r$. We plot the value of ${\cal T}(r)$ for each realization 
as a blue point. The largest computed escape times (for $r=4L$)
are $38,506\,$ps for K$^+$ and $47,212\,$ps for Na$^+$. They are 
outside the range of panels in Figure~\ref{figure5}, but the 
majority of data poins are included in this figure.  
We also plot average $\langle {\cal T}(r) \rangle$ (red solid line) 
together with 95\% confidence intervals. They are compared with 
theoretical results obtained for the BD model~(\ref{BDSDE}).
The escape time distribution for the BD model~(\ref{BDSDE}) has mean
equal to $\langle {\cal T}(r) \rangle = L^2/(6D)$ and standard deviation
$L^2/(3 \sqrt{10} \, D).$ The corresponding theoretical 95\% confidence 
interval (for 100 samples) is
\begin{equation}
\left( 
\frac{L^2}{6D} - 1.96 \frac{L^2}{30 D}, \,
\frac{L^2}{6D} + 1.96 \frac{L^2}{30 D}
\right).
\label{theorconfinterval}
\end{equation}
This interval is visualized as the green area in Figure~\ref{figure5}.
We note that it would be relatively straightforward to continue the
presented multiscale computation and simulate ion diffusion in domains 
covering the whole cell. The most computationally intensive part
is all-atom MD simulation in $\Omega_1 \cup \Omega_2$. However, once
the ion enters $\Omega_5$, we can compute its trajectory very efficiently. 
We could further increase the BD time step in parts of $\Omega_5$ which
are far away from $\Omega_4$, or we could use event-based algorithms,
like Green's-function reaction dynamics~\cite{vanZon:2005:GFR} or 
First-passage kinetic Monte Carlo method~\cite{Opplestrup:2009:FKM},
to compute the ion trajectory in region $\Omega_5$.

\section{Discussion}
\label{secdiscussion}

In this paper, we have introduced and studied the coarse-grained
model~(\ref{BDXeqAAA})--(\ref{BDZeqAAA}) of an ion in aqueous 
solution. We have parameterized this model using all-atom MD 
simulations for four ions (K$^+$, Na$^+$, Ca$^{2+}$ and Cl$^-$)
and showed that this model provides an intermediate description
between all-atom MD and BD simulations. It can be used both with 
MD time step $\Delta t$ (to couple it with all-atom MD simulations)
and BD time step $\Delta T$ (to couple it with BD description~(\ref{BDSDE})).
In particular, the coarse-grained model enables multiscale simulations
which use all-atom MD and BD in different parts of the computational
domain. 

In Section~\ref{seccoupleMDBD}, we have illustrated this
multiscale methodology using a first passage type problem where 
we have reported the time taken by an ion to reach a specific 
distance. Possible applications of this multiscale methodogy
include problems where a modeller considers all-atom MD in several
different parts of the cell (for example, close to binding sites 
or ion channels) and wants to use efficient BD simulations to 
transport ions by diffusion between regions where MD is used.
The proposed approach thus enables the inclusion of MD-level of detail
in computational domains which are much larger than would
be possible to study by direct MD simulations.

Although the illustrative simulations in Section~\ref{seccoupleMDBD}
are reported over distances of the order of $10^2$ \AA, this is not 
a restriction of the method. Most of the computational time is spent 
by considering all-atom MD in $\Omega_1 \cup \Omega_2$. BD uses much 
larger time step which enables us to futher extend BD region $\Omega_5$ 
(and consequently, the original domain $\Omega$). Moreover, if we 
are far away from MD domain $\Omega_1$, we can further increase the 
efficiency of BD simulations by using different BD time steps in 
different parts of the BD subdomain $\Omega_5$~\cite{Erban:2014:MDB}, 
or by using event-based BD
algorithms~\cite{vanZon:2005:GFR,Opplestrup:2009:FKM}.
The computational intensity of BD simulations can be further 
decreased by using multiscale methods which efficiently and accurately
combine BD models with lattice-based (compartment-based) models
\cite{Flegg:2012:TRM,Robinson:2015:MRS}. Such a strategy have been
previously used for modelling intracellular calcium dynamics \cite{Dobramysl:2015:PMM,Flegg:2013:DSN} or actin dynamics in 
filopodia~\cite{Erban:2013:MSR}, and enables us to extend 
both temporal and spatial extent of the simulation. 

In the literature, MD methods have been used to estimate parameters
of BD simulations of ions~\cite{Allen:2000:MDE}. There has also been
a lot of progress in systematic coarse-graining of 
MD simulations~\cite{Saunders:2013:CGM}. The approach presented 
in this paper not only uses all-atom MD simulations to estimate 
parameters of a coarser description, but it also designs a multiscale 
approach where both methods are used during the same simulation. 
Methods which adaptively change the resolution of MD on demand have 
been previously reported in~\cite{Praprotnik:2008:MSS,Nielsen:2010:RPA}.
They include algorithms which couple all-atom MD with coarse-grained
MD. The coarse-grained model developed in this work does not include
any water molecules and has different application areas. One of 
them is modelling of calcium induced calcium release through IP$_3$R
channels~\cite{Dobramysl:2015:PMM} which is discussed as a motivating 
example in Introduction. MD simulations in this paper use the
three-site SPC/E model of water. An open question is to extend our 
observations and analysis to other MD models of water, which include 
both more detailed water models with additional
sites~\cite{Huggins:2012:CLW,Mark:2001:SDT}
and coarse-grained MD models of water~\cite{Praprotnik:2007:ARS}.

\section*{Acknowledgements}

I would like to thank the Royal Society for a University Research 
Fellowship and the Leverhulme Trust for a Philip Leverhulme Prize.  

\newpage

{
\small
\bibliographystyle{unsrt}
\setlength{\bibsep}{1.9pt}

}

\end{document}